\documentclass[floatfix,pra,twocolumn,superscriptaddress]{revtex4-2}
\usepackage[utf8]{inputenc}
\usepackage[american]{babel}
\usepackage[pdftex]{graphicx}  
\usepackage{graphicx, xcolor}
\usepackage{dcolumn}
\usepackage{bm}
\usepackage{dsfont}
\usepackage{amsmath,amsthm,amssymb}
\usepackage{hyperref}
\usepackage{amsthm}
\usepackage{xcolor}
\usepackage{braket}
\hypersetup{colorlinks,bookmarksopen,bookmarksnumbered,
citecolor=red,
linkcolor=red,
pdfstartview=false,
urlcolor=red}
\usepackage{graphicx}
\newtheorem{thm}{Theorem}
\newtheorem{cor}{Corollary}

\begin{document}
\title{Measuring Nonstabilizerness via Multifractal Flatness}

\author{Xhek Turkeshi}
\affiliation{JEIP, USR 3573 CNRS, Coll\`{e}ge de France, PSL Research University, 11 Place Marcelin Berthelot, 75321 Paris Cedex 05, France}
\author{Marco Schir\`o}
\affiliation{JEIP, USR 3573 CNRS, Coll\`{e}ge de France, PSL Research University, 11 Place Marcelin Berthelot, 75321 Paris Cedex 05, France}
\author{Piotr Sierant}
\affiliation{ICFO-Institut de Ci\`encies Fot\`oniques, The Barcelona Institute of Science and Technology, Av. Carl Friedrich Gauss 3, 08860 Castelldefels (Barcelona), Spain}
\date{\today}

\begin{abstract}
Universal quantum computing requires non-stabilizer (magic) quantum states. Quantifying the nonstabilizerness and relating it to other quantum resources is vital for characterizing the complexity of quantum many-body systems. In this work, we prove that a quantum state is a stabilizer if and only if all states belonging to its Clifford orbit have a flat probability distribution on the computational basis. This implies, in particular, that multifractal states are non-stabilizer. We introduce multifractal flatness, a measure based on the participation entropy that quantifies the wave function distribution flatness. 
We demonstrate that this quantity is analytically related to the stabilizer entropy of the state and present several examples elucidating the relationship between multifractality and nonstabilizerness. In particular, we show that
the multifractal flatness provides an experimentally and computationally viable nonstabilizerness certification. Our work unravels a direct relation between the nonstabilizerness of a quantum state and its wave function structure.
\end{abstract}
\maketitle

\section{Introduction}
Originally developed in the context of fault-tolerant quantum computation and error-correction~\cite{gottesman1998theoryoffaulttolerant,eastin2009restrictions}, stabilizer states and operations play an essential role in quantum information theory~\cite{nielsen00}, in our understanding of thermalization \cite{Fritzsch21, Claeys21} and entanglement propagation in many-body systems~\cite{nahum2017quantum,nahum2018operator,Fisher2023}, or providing insights on the holography~\cite{Pastawski2015}. 
Yet, their classical simulability~\cite{aaronson2004improvedsimulationof} implies that additional ingredients are required to attain computational quantum speed-up~\cite{shor1997polynomialtimealgorithmsfor,feynman1982simulatingphysicswith,deutsch1985quantumtheory}.
Quantifying the nonstabilizerness (or ``magic'') of a quantum state is, therefore, a key task in the resource theory of 
quantum computation~\cite{Kitaev2003, Gottesman1999demonstratingtheviability,bravyi2005universalquantumcomputation, Veitch2014theresourcetheory,bravyi2016tradingclassicaland,chitambar2019quantumresourcetheories,kelly2023coherence,weinstein2022scrambling},
related to the classical computational cost of simulation quantum physics~\cite{Howard2014,bravyi2016tradingclassicaland,seddon2021quantifying,Koukoulekidis2022fasterborn} and to onset of quantum chaos~\cite{haferkamp2022randomquantum,leone2021quantumchaosis}.
While various nonstabilizerness monotones have been proposed~\cite{campbell2011catalysisandactivation,Veitch2014theresourcetheory,howard2017application,Wang2019,Beverland2020,liu2022manybodyquantummagic,bu2022complexity,jiang2023lowerboundfor}, their geometrical nature requires minimization over large spaces 
resulting in prohibitive computational costs already for few-qubit systems.
In fact, quantifying nonstabilizerness in many qubit systems remains a major challenge.
Stabilizer entropy provides a (generally non-monotone~\cite{haug2023stabilizerentropiesand}) measure of nonstabilizerness that is calculable for few-qubit states~\cite{leone2022stabilizerrenyientropy,leone2023phase,oliviero2021transitionsinentanglementcomplexity,oliviero2022magicstateresourcetheory} and that is related to the flatness of the entanglement spectrum~\cite{tirrito2023quantifying}. 
While being experimentally measurable in present quantum devices via randomized measurement protocols or Bell measurements~\cite{oliviero2022measuringmagicon,haug2023scalabremeasuresof}, recent works demonstrated that the stabilizer entropy is efficiently computable for many-body matrix product states~\cite{haug2023quantifyingnonstabilizernessof,haug2023stabilizerentropiesand,lami2023quantum}. 
These results allow for studying static and dynamical properties of nonstabilizerness in one-dimensional many-body systems. 

A seemingly unrelated quantity is the inverse participation ratio and the participation entropy of many-body wave functions. Originally developed in the context of Anderson localization~\cite{castellani1986multifractal,evers2000fluctuationsofthe,evers2008andersontransitions,rodriguez2009,rodriguez2010critical,DeLuca2013,sierant2023universality}, these quantities measure the spread of the many-body states on a certain basis of the Hilbert space. Studies in ground states of many-body systems~\cite{stephan2009,stephan2010,stephan2014shannon,luitz2014universalbehaviorbeyond,fradkin2006entanglemententropyof,zaletel2011logarithmicterms,alcaraz2013universalbehaviorof,lindinger2019manybodymultifractality,luitz2014shannonrenyientropy,atas2012multifractality,Luitz2014}, in the problem of many-body localization~\cite{luitz2020multifractality,mace2019multifractal,tarzia2020manybody,solorzano2021,Pietracaprina2021,rodriguez2009,Monthus2016,pausch2021chaosanderdoficity,sierant2023stability,detomasi2020multifractality,roy2021fockspace,detomasi2021rare,biroli2021outofequilibrium,roy2022hilbertspace}, and monitored unitary dynamics~\cite{sierant2022universalbehaviorbeyond, Sierant22}, demonstrated that most quantum states have multifractal features~\cite{stanley1988multifractalphenomenain}, i.e., they are described by different fractal properties in different geometrical regions. 

This work demonstrates that nonstabilizerness is directly encoded in the structure of wave functions. 
Employing methods inspired by recent works~\cite{tirrito2023quantifying,carrasco2022entanglement}, we show that a state has a flat participation distribution along its Clifford orbit if and only if it is a stabilizer state. Conversely, a state with non-flat participation distribution, for instance, a multifractal state, has non-stabilizerness.
After reviewing the key concepts of interest, we introduce a \textit{multifractal flatness}, a measure of 
the flatness of participation distributions that we prove is a simple function of the stabilizer entropy.  This allows us to demonstrate that the participation entropy in the computational basis provides a useful magic witness that is amenable to computational methods. 
We illustrate the relation between nonstabilizerness and participation entropy on examples of a single qubit, many-body product states, and Haar random states. 
Finally, we demonstrate that the nonstabilizerness quantifier introduced in this manuscript is measurable in noisy intermediate-scale quantum devices~\cite{ferris2022quantum,Preskill2018quantumcomputingin,fraxanet2022coming}. 

\section{Stabilizer and participation entropies} 
We consider a system of $N$ qubits with Hilbert space dimension $d=2^N$, and denote $\{\sigma^\alpha\}_{\alpha=0,1,2,3}$ the Pauli matrices ($\sigma^0=\openone$), $|0\rangle,|1\rangle$ the local computational basis of $\sigma^3$ and $\mathcal{P}_N$ the set of all $N$-qubit Pauli strings. The Clifford group $\mathcal{C}_N$ is the subset of unitary operations that map a Pauli string into a \emph{single} Pauli string.
The stabilizer entropy is defined for a pure normalized state $|\Psi\rangle$ as~\cite{leone2022stabilizerrenyientropy}
\begin{equation}
    M_q(|\Psi\rangle) = \frac{1}{1-q} \log_2 \sum_{P\in \mathcal{P}_N} \frac{(\langle \Psi|P |\Psi\rangle)^{2q}}{d}.\label{eq:se}
\end{equation} 
It is a measure of nonstabilizerness as: (i) $M_q(|\Psi\rangle)\ge 0$, with the equality holding if and only if the state is a stabilizer $|\Psi\rangle \in \mathrm{STAB}$, (ii) it is invariant under Clifford conjugation $M_q(C|\Psi\rangle) = M_q(|\Psi\rangle)$ for any $C\in \mathcal{C}_N$, (iii) it is additive $M_q(|\Psi\rangle\otimes|\Phi\rangle) = M_q(|\Psi\rangle) + M_q(|\Phi\rangle)$~\cite{leone2022stabilizerrenyientropy}. 
Despite the fact that the stabilizer entropy is not a magic monotone for generic R\'enyi index $q$~\cite{haug2023stabilizerentropiesand},
it is computationally amenable compared to the magic robustness and stabilizer fidelity~\cite{Howard2014}. 
We aim to show that the stabilizer entropy $M_2(|\Psi\rangle)$ is related to the structure of the wave function quantified by the participation entropy.

Given a pure state $|\Psi\rangle$, we introduce the participation distribution in the computational basis $\mathcal{B}\equiv \{|\vec{\sigma}\rangle,\ \sigma_i=0,1,\ i=1,\dots,N\}$ as the probability distribution $p(\vec{\sigma})\equiv |\langle \vec{\sigma}|\Psi\rangle|^2$. Then, the participation entropy is 
\begin{equation}
    S_q( \ket{\Psi} ) = (1-q)^{-1} \log_2 I_q(|\Psi\rangle),
    \label{eq:pe}
\end{equation}
where ${I_q(|\Psi\rangle) = \sum_{ \vec{\sigma} \in \mathcal{B}} p(\vec{\sigma})^{q}}$ is the inverse participation ratio. The participation entropy~\eqref{eq:pe} quantifies the spreading of the state $|\Psi\rangle$ over the basis $\mathcal{B}$. Conventionally, in condensed matter settings, the system size dependence of the participation entropy is parametrized as~\cite{backer2019multifractal, mace2019multifractal, sierant2023universality}
\begin{equation}
    S_q = D_q N + c_q,\label{eq:syssize}
\end{equation}
where $D_q$ is the multifractal dimension and $c_q$ is a sub-leading term. 
In the field of Anderson and many-body localization transition~\cite{evers2008andersontransitions}, it is customary to denote the state $|\Psi\rangle$ localized (fully extended) when $D_q=0$ ($D_q=1$). The intermediate regimes, for which $0<D_q<1$ and $D_q$ depend non-trivially on the R\'enyi index $q$ are said to be multifractal~\footnote{It is also possible that $0<D_q<1$ and $D_{q_1}=D_{q_2}$ for all $q_1,q_2>0$. Such a state is called fractal.}. 
Here we are concerned with the \textit{participation flatness}, occurring when $S_{q_1}( \ket{\Psi} )= S_{q_2}( \ket{\Psi} )$ for all $q_1,q_2>0$. (This condition is equivalent to $p(\vec{\sigma})$ being uniform in its domain, hence justifying the name). 
Participation flatness implies the absence of wave function multifractality. 
However, we remark that the two notions are not equivalent: participation flatness is defined for a given state at a fixed system size $N$. Instead, the multifractality is intrinsically related to the $N$ dependence of the participation entropy~\eqref{eq:syssize}. In particular, as we will discuss in the following, a state can be fully-extended ($D_q=1$) or localized ($D_q=0$) and still not have a flat participation distribution due to the non-trivial $q$-dependence of $c_q$.

Finally, let us note that the stabilizer entropy~\eqref{eq:se} has an intrinsic basis dependence in the choice of Pauli strings as generators of the operator space. This basis dependence is a feature of the nonstabilizerness and different frames yield different results~\cite{howard2017application}. Similarly, the participation entropy~\eqref{eq:pe}, and hence also the participation flatness, depends explicitly on the choice of many-body basis $\mathcal{B}$. 
Therefore, we choose the Pauli strings to quantify the degree of nonstabilizerness and consider the computational basis to probe the many-body wave function structure.

\section{Nonstabilizerness and multifractality}
Stabilizer states are not multifractal and always possess a flat participation distribution, as shown in~\cite{sierant2022universalbehaviorbeyond}, which also provides an efficient way to compute participation entropy of stabilizer state using its tableaux representation~\cite{aaronson2004improvedsimulationof, gidney2021stimfaststabilizer}. As we argue in the following, the converse statement is also true, leading to a novel characterization of nonstabilizerness in terms of the participation entropies of the many-body wave function. 

\begin{figure}
    \centering
    \includegraphics[width=\columnwidth]{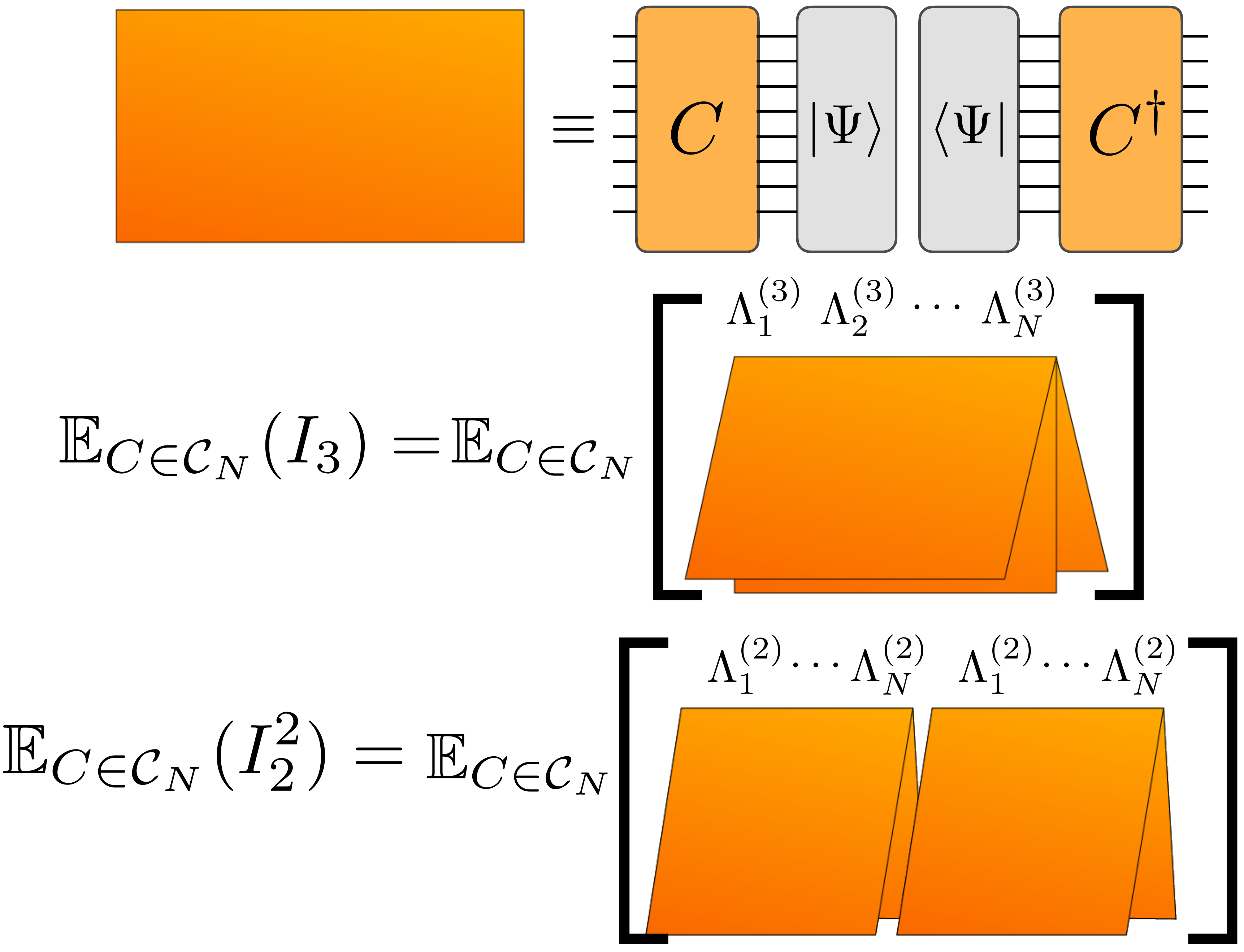}
    \caption{Pictorial representation of the replica picture. Each \emph{page} is a folding of $C|\Psi\rangle$  and its adjoint. Applying $\Lambda^{(q)}_k$ on each site imply a sum over the same physical index and results in \emph{book} boundary conditions.  }
    \label{fig:replica}
\end{figure}

Before stating the main results of this work, we introduce the \emph{multifractal flatness} $\mathcal{F}(|\Psi\rangle) = I_3(|\Psi\rangle) - I_2^2(|\Psi\rangle)$, which is a measure of the participation flatness. 
By concavity of the participation entropy and using Jensen's inequality, it is easy to see that $\mathcal{F}(|\Psi\rangle) \ge 0$, with the equality holding if and only if the participation distribution $p({\vec{\sigma}})$ is flat or, equivalently, when $S_{q_1}( \ket{\Psi} )= S_{q_2}( \ket{\Psi} )$ for all $q_1,q_2>0$.
\begin{thm} 
The average of the multifractal flatness $\mathcal{F}$ over the Clifford orbit $\mathcal{C}_\Psi\equiv \{ C|\Psi\rangle \ | C \in \mathcal C_N\}$ is a measure of nonstabilizerness, with
\begin{equation}
    \overline{\mathcal{F}}(|\Psi\rangle)\equiv \mathbb{E}_{C\in \mathcal{C}_N}[\mathcal{F}(C|\Psi\rangle)] = \frac{2(1-2^{-M_2(|\Psi\rangle)})}{(d+1)(d+2)}.\label{eq:thm}
\end{equation}
\end{thm}
Let us highlight an immediate consequence that conceptually bridges the participation flatness to that of nonstabilizerness.
\begin{cor} A state is a stabilizer if and only if every element of its Clifford orbit is participation flat. Vice versa, a state is non-stabilizer if and only if it exists $C\in \mathcal{C}_N$ for which $C|\Psi\rangle$ is not participation flat. 
\end{cor}

As already mentioned, a stabilizer state has flat participation distribution~\cite{sierant2022universalbehaviorbeyond}. In this case, Eq.~\eqref{eq:thm} is trivially satisfied.
Conversely, we must prove the flatness of the participation distribution for all the states in $\mathcal{C}_{{\Psi}}$, or in other words, that there is no $C\in \mathcal{C}_N$ for which $S_q(C|\Psi\rangle)$ is $q$ dependent. From the positive monotonicity of $\mathcal{F}$, this fact is equivalent in proving Eq.~\eqref{eq:thm} holds for generic states. 
To this end, let us rephrase $I_q^r$ in the replica formalism as
\begin{equation}
    I_q^r = \mathrm{tr}\left[| \Phi^{(rq)}\rangle\langle \Phi^{(rq)}| \left(\Lambda_1^{(q)}\right)^{\otimes r}\cdots \left(\Lambda_N^{(q)}\right)^{\otimes r} \right],\label{eq:start}
\end{equation}
where $ |\Phi^{(rq)}\rangle=|\Psi\rangle^{\otimes rq}$ is the replica state and $\Lambda^{(q)}_k = (|0\rangle\langle 0|)^q + (|1\rangle\langle 1|)^q$ are the operators enforcing the book replica boundary condition~\cite{stephan2010,sierant2022universalbehaviorbeyond}, cf. Fig.~\ref{fig:replica}.
From Eq.~\eqref{eq:start} we see that $I_q^r$ has a permutation invariance $\mathcal{S}_q^{\otimes r}$ over the replica space. This fact will allow for simplifications in evaluating the orbit average in Eq.~\eqref{eq:thm}. First, let us compute the $q=3, r=1$ term. Using the 3-design property of the Clifford group~\cite{zhu2017multiqubitcliffordgroups,webb2016clifford} we have 
\begin{equation}
    \mathbb{E}_{\mathcal{C}_N}[C^{\otimes 3}|\Phi^{(3)}\rangle\langle \Phi^{(3)}|(C^\dagger)^{\otimes 3}] = \frac{6}{d(d+1)(d+2)}\Pi_{[3]},
\end{equation}
with $\Pi_{[k]} = \sum_{\pi\in \mathcal{S}_k} U_{\pi}/k!$ the projector onto the symmetric permutation of $k$ elements. It follows that $\mathbb{E}_{C\in\mathcal{C}_N}[I_3(C|\Psi\rangle)] = {6}/{((d+1)(d+2))}$. 
Instead, the $q=2,r=2$ require the average of the $|\Phi^{(4)}\rangle$. This object is less trivial and needs insights into the commutant of the Clifford group, cf. Ref.~\cite{zhu2016thecliffordgroup,nezami2020multipartiteentanglementin}.
We have
\begin{equation}
    \mathbb{E}_{\mathcal{C}_N}[C^{\otimes 4}|\Phi^{(4)}\rangle\langle \Phi^{(\otimes 4)}|(C^\dagger)^{4}] = \beta_+(|\Psi\rangle) \Pi_+ + \beta_-(|\Psi\rangle) \Pi_-,
\end{equation}
with $\Pi_+ = \Pi_{N,4} \Pi_{[4]}$, $\Pi_- = (1-\Pi_{N,4}) \Pi_{[4]}$ two projectors and $\Pi_{N,4} = \sum_{P\in \mathcal{P}_N} P^{\otimes 4}/d^2$, while the coefficients are defined as $\beta_+ = 6||\Xi(|\Psi\rangle)||^2_2/((d+1)(d+2))$ and $\beta_- = 24(1-||\Xi(|\Psi\rangle)||^2_2)/((d^2-1)(d+2)(d+4))$, with $||\Xi(|\Psi\rangle)||^2_2 = \sum_{P \in \mathcal P_N} \bra{\Psi} P \ket{\Psi}^4/d^2$. 
A simple computation imposing the book boundary condition in the replica space~\cite{sierant2022universalbehaviorbeyond} gives
\begin{equation}
    \mathbb{E}_{C\in\mathcal{C}_N}[I_2^2(C|\Psi\rangle)] = \frac{4-2d ||\Xi(|\Psi\rangle)||^2_2}{(d+1)(d+2)}.
\end{equation}
Combining these results, it follows Eq.~\eqref{eq:thm} and the proof's conclusion for $N\ge 3$. The case $N=1,2$ have a rank deficient Clifford group commutant~\cite{montealegremora2022dualitytheoryfor,MontealegreMora2021} and must be evaluated separately. Nevertheless, one can show that the final result Eq.~\eqref{eq:thm} holds, recalling $d=2^N$. $\square$

A few remarks are in order here. At a practical level, the right-hand side of Eq.~\eqref{eq:thm} can be evaluated via Monte Carlo sampling over the Clifford group $\mathcal{C}_N$, i.e., $\overline{\mathcal{F}}_\mathrm{mc}\equiv \sum_{C} \mathcal{F}(C|\Psi\rangle)/\mathcal N_{\mathrm{real} }$ with $\mathcal N_{\mathrm{real} }$ the number of random choices of Clifford unitaries $C$. We recall that random drawing from the Clifford group is efficiently implementable~\cite{Bravyi_2021}, and we present an explicit example of the Monte Carlo estimation in the next sections. 

While our theorem~\eqref{eq:thm} links the multifractal flatness averaged over the whole Clifford orbit $\mathcal C_{\Psi}$ to the stabilizer entropy $M_2$, we remark that $\mathcal{F}(C^*|\Psi\rangle)$ for a fixed $C^*\in \mathcal{C}_N$ is a convenient witness of nonstabilizerness, being computationally cheap compared to~\eqref{eq:se}. Indeed, $\mathcal{F}(C|\Psi\rangle)\geq 0$ for any $C$ and $\ket{\Psi}$ and is non-zero only if the state it not a stabilizer. Hence, if $\mathcal{F}(C^*|\Psi\rangle)>0$ we certify that the state has some amount of nonstabilizerness.

Additionally, the multifractal flatness $\mathcal{F}$ can be extended to the class $\mathcal{F}(|\Psi\rangle;q,m) = I_q(|\Psi\rangle) - I_{(k-1+q)/m}^m(|\Psi\rangle)$, with $m,q>0$. While we conjecture a result similar to~\eqref{eq:thm} holds, the rapidly growing commutant dimension of the Clifford group hinders analytical insights already for $q,k = 5$~\cite{Gross2021}. Nevertheless, these terms would scale $O(d^{-3})$, therefore being more difficult to resolve than~\eqref{eq:thm} for practical purposes. 
Lastly, we note an expression similar to~\eqref{eq:thm} resembles analogous results for the entanglement spectrum~\cite{tirrito2023quantifying}, where an additional dependence on the entanglement bipartition is present.

In the remaining, we illustrate the relation between nonstabilizerness and the multifractal flatness $\mathcal F$ for several examples of quantum states. In particular, we show that resolving $\overline {\mathcal{F} }$ generally requires exponential resources in system size for generic systems, as shown in a numerical example below. Yet, we show that $\overline{\mathcal{F}}$ can be estimated in current noisy intermediate-scale quantum devices, provided the fidelity is preserved. 

\section{Multifractality and nonstabilizerness: examples}
\subsection{Single qubit}
Let us start with the intuitive example of a single qubit ($N=1$) and highlight the relationship between stabilizer and participation entropies of the 
quantum state
\begin{equation}
\label{eq:singleQubit}
    |\Psi_1\rangle = 
    \cos\left(\frac{\theta}{2}\right)|0\rangle + e^{i \phi} \sin\left(\frac{\theta}{2}\right)|1\rangle, 
\end{equation}
where $\theta, \phi$ are real parameters.
The stabilizer entropy \eqref{eq:se} of this state is given by
\begin{equation}
\begin{split}
    M_q(|\Psi_1\rangle) &= \frac{1}{1-q}\log_2\left(\frac{1+\cos ^{2 q}(\theta )+\Omega(\theta,\phi)}{2}\right)\\
    \Omega(\theta,\phi) &= (\sin (\theta ) \sin (\phi ))^{2 q}+(\sin (\theta ) \cos (\phi ))^{2 q}
\end{split}
\label{eq:N1se}
\end{equation}
which is non-vanishing for a generic choice of $\theta$ and $\phi$.  Calculation of the participation entropy~\eqref{eq:pe} yields $S_q( \ket{\Psi_1}) = \log_2[\cos^{2q}(\theta/2)+\sin^{2q}(\theta/2)]/(1-q)$,
which depends non-trivially on $q$ for a generic parameter choice, providing an example of a typical scenario in which a non-stabilizer state does not have a flat participation distribution. However, for a fine-tuned choice  $\theta = \pi/2$, the participation entropy is equal to unity, independently of the value of $q$, hence $\mathcal F (\ket{\Psi_1})=0$. Still, the state $\ket{\Psi_1}$ has a nontrivial nonstabilizerness for  $\theta = \pi/2$ and a generic value of $\phi$. This shows that flatness of participation distribution in a selected point of the Clifford orbit $\mathcal C_{\Psi_1}$
does not imply that the state is a stabilizer state, illustrating the importance of the average over the whole Clifford orbit in \eqref{eq:thm}. Indeed, for a single choice of $C\in \mathcal{C}_N$, $\mathcal{F}$ is only a magic witness: if it is non-zero, we know that the state has nonstabilizerness, but the converse is not true.
If we act with a Hadamard gate $H \in \mathcal C_1$ on the state $\ket{\Psi_1}$, we discover a nontrivial $q$ dependency 
$S_q(   H \ket{\Psi_1} ) = (\log_2[((1+\sin\theta\cos\phi)/2)^{q}+((1-\sin\theta\cos\phi)/2)^{q}])/(1-q)$ for generic values of $\phi$ even at  $\theta = \pi/2$, consistently with the nonstabilizerness of the state $\ket{\Psi_1}$ revealed by the non-vanishing value of $M_q(\Psi)$ \eqref{eq:N1se}. Its multifractal flatness is $\mathcal{F}(H|\Psi\rangle)=(\sin ^2(\theta ) \cos ^2(\phi )-\sin ^4(\theta ) \cos ^4(\phi ))/4$, that must be compared with the Clifford orbit average in Eq.~\eqref{eq:thm}, given by  $\overline{\mathcal{F}}(|\Psi\rangle)= \sin ^2(\theta ) \left(-2 \sin ^2(\theta ) \cos (4 \phi )+7 \cos (2 \theta )+9\right)/96$.

\subsection{Many-qubit product states}
We consider now a state $\ket{\Psi_N}$ of $N$ qubits which is a product state of the single qubit states $\ket{\Psi_1}$ \eqref{eq:singleQubit}, i.e., $\ket{\Psi_N} = \ket{\Psi_1}^{\otimes N}$. This state is not entangled but has an extensive stabilizer entropy $M_q( \ket{\Psi_N}) = N M_q( \ket{\Psi_1})$, where $M_q( \ket{\Psi_1})$ is given by \eqref{eq:N1se}, i.e. $\ket{\Psi_N}$ is a magic state. Moreover, it is easy to verify that $S_q( \ket{\Psi_N}) = N S_q( \ket{\Psi_1})$, i.e. for a generic choice of $\theta$ and $\phi$, the state $M_q( \ket{\Psi_N})$ is multifractal with the multifractal dimension $D_q = S_q( \ket{\Psi_1})$ and the sub-leading term $c_q =0$. This is an example of a situation in which the presence of multifractality at a given point of the stabilizer orbit of $\ket{\Psi_N}$ implies that the state is non-stabilizer $\ket{\Psi_N}$. However, this is not the case for the fine-tuned choice $\theta = \pi / 2$, for which the multifractality flatness vanishes, $\mathcal F(\ket{\Psi_N}) = 0$, and the state $\ket{\Psi_N}$ is fully extended, $D_q=1$, while still being non-stabilizer for a generic value of $\phi$. Similarly as for the single qubit, the action of the Hadamard gates on $\ket{\Psi_N}$ yields a state with participation distribution that is not flat -- with the multifractal flatness depending on the number of Hadamard gates considered, thus revealing the nonstabilizerness of the state.

\subsection{Random Haar states}
\label{sec:random}
We consider a random state $|\Psi\rangle$ and quantify its nonstabilizerness by calculating the stabilizer entropy via~\eqref{eq:thm}. Any such state is obtainable for a (random) Haar unitary $U\in\mathcal{U}(d)$ acting on a reference state $|\Psi_0\rangle$. For convenience and without loss of generality, $|\Psi_0\rangle = |0\rangle^{\otimes N}$.
We begin by computing the Haar average $\mathbb{E}_{U\in \mathcal{U}(d)}\left[ \overline{\mathcal{F}[U|\Psi\rangle]}\right]$. As we shall argue, in the limit of large systems $N\gg 1$, $\overline{\mathcal{F}}$ is self-averaging for any fixed realization $|\Psi\rangle = U|\Psi_0\rangle$.

First, let us note that $\mathbb{E}_{U\in \mathcal{U}(d)}[\overline{\mathcal{F}}[U|\Psi_0\rangle]] = \mathbb{E}_{U\in \mathcal{U}(d)}\mathcal{F}[U|\Psi_0\rangle]$.
This fact follows from the Clifford group being a subgroup of the unitary ensemble $\mathcal{C}_N\subset \mathcal{U}(d)$ and by the unitary invariance of the Haar measure. 
Recalling that  $\mathbb{E}_{U\in \mathcal{U}(d)}\left[\left(U|\Psi_0\rangle\langle \Psi_0|U^\dagger\right)^{\otimes k}\right] = k!\Pi_{[k]}/(d(d+1)\dots (d+k-1))$ and using the book boundary conditions in replica space~(cf. Fig.~\ref{fig:replica}), it follows that
\begin{equation}
\begin{split}
    \mathbb{E}_{U\in\mathcal{U}(d)}&\left[I_q^r(U|\Psi_0\rangle)\right] =\\
    &\frac{\sum_{\lambda  \vdash r} d(d-1)\dots (d-K_\lambda +1)((\lambda_k q)!)^{n_k} a_{\lambda }}{d(d+1)\dots (d+rq-1)}.
\end{split}
     \label{eq:res}
\end{equation}
In the above formula Eq.~\eqref{eq:res}, $\lambda=\{(\lambda_k,n_k)\}_{k=1,\dots,K_\lambda}$ runs over the partitions of $r=\sum_{k=1}^{K_\lambda} n_k \lambda_k $, 
and the remaining coefficients are $a_\lambda$ are given by 
\begin{equation}
    a_\lambda = \frac{r!}{\left[\prod_{k=1}^{K_\lambda} \left(\lambda_k!\right)^{n_k}\right] \left[\prod_{k=1}^{K_\lambda} (n_k!)\right] }.
\end{equation}
Specializing to $q=3,r=1$ and $q=r=2$ we have
\begin{equation}
    \overline{\mathcal{F}}^U\equiv \mathbb{E}_{U\in \mathcal{U}(d)}[\overline{\mathcal{F}}[U|\Psi_0\rangle]] = \frac{2(d-1)}{(d+1)(d+2)(d+3)}. \label{eq:result}
\end{equation}
In particular, we recover the stabilizer entropy computation in Ref.~\cite{leone2022stabilizerrenyientropy}, with $\mathbb{E}_{U\in \mathcal{U}(d)}\left[2^{-M_2[U|\Psi_0\rangle]}\right] =4/(d+3)$. 
For a realization $U\in \mathcal{U}(d)$, the state $|\Psi\rangle = U|\Psi_0\rangle$ has $\overline{\mathcal{F}}\simeq \overline{\mathcal{F}}^U$, with an exponentially small error in system size $N$ as a consequence of quantum typicality~\cite{Goold2016,Wilde2013}. 
Indeed, the standard deviation of $\overline{\mathcal{F}}$ over the Haar ensemble is at leading order in $1/d$ is given by $\mathrm{std}_{U\in \mathcal{U}(d)}(\overline{\mathcal{F}}[U|\Psi_0\rangle])\simeq 2\sqrt{34}/d^{5/2} + O\left(d^{-6}\right)$~\footnote{The exact expression is 
\begin{align*}
  &  \mathrm{std}_{U\in \mathcal{U}(d)}(\overline{\mathcal{F}}[U|\Psi_0\rangle])^2 = \\
  &\frac{8 \left(17 d^5+42 d^4-106 d^3-72 d^2+449 d-330\right)}{(d+1)^2 (d+2)^2 (d+3)^2 (d+4) (d+5) (d+6) (d+7)}.
\end{align*}
This formula requires evaluating generic correlators $\mathbb{E}_{U\in\mathcal{U}(d)}I_{q_1}^{r_1}[U|\Psi_0\rangle]I_{q_2}^{r_2}[U|\Psi_0\rangle]$ and result in expression similar to Eq.~\eqref{eq:res} but more involved. 
}. 

\begin{figure}[t]
    \centering
    \includegraphics[width=\columnwidth]{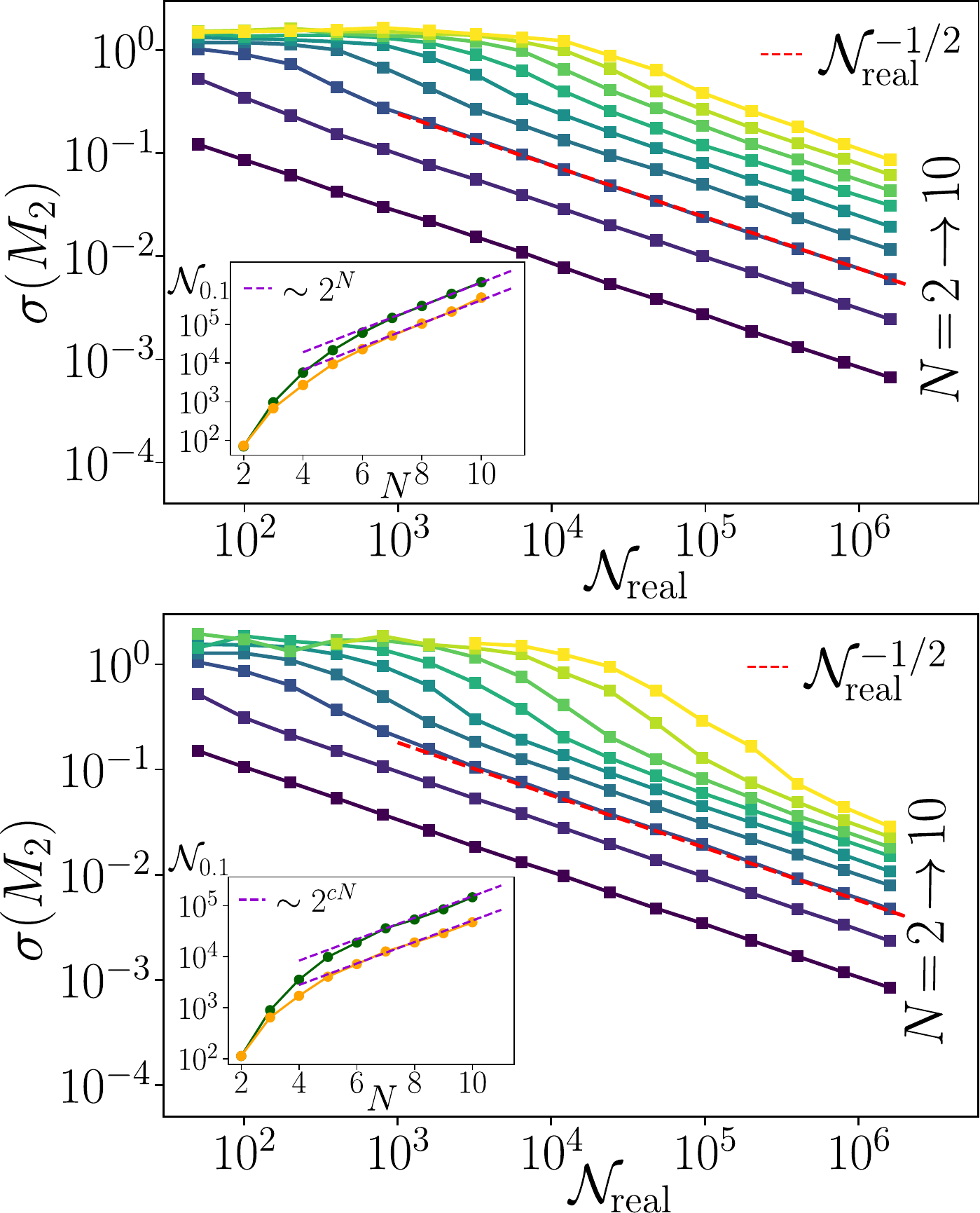}
    \caption{Determining stabilizer entropy with multifractal flatness $\mathcal{F}$. Upper panel: statistical error $\sigma(M_2)$ of stabilizer entropy determined from $\mathcal N_\mathrm{real}$ measurements of $\mathcal{F}$ for a random Haar state; the asymptotic $\mathcal {N}^{-1/2}_\mathrm{real}$ scaling is denoted by a dashed line; data (color online) shown for number of qubits $N=2,3,\ldots,10$; the green line in the inset shows the number of measurements $\mathcal{N}_{0.1}$ needed to achieve error $\sigma(M_2)=0.1$ as function of $N$, the orange line shows $\mathcal{N}_{0.1}$ for a protocol in which a full layer of 2-body Clifford gates acts between each measurement of $\mathcal{F}$. Bottom panel: the same, but for the product state $\ket{\Psi_N}$ with $\phi=\pi/2$ and $\theta = \pi/2$; the observed scaling $\mathcal{N}_{0.1} \sim 2^{cN}$ is slower than for random Haar state (since $c\approx 0.7$), but still exponential in $N$.}
    \label{fig:ex1}
\end{figure}

The random Haar states constitute an ensemble of nonstabilizer states with stabilizer entropy $M_2[U|\Psi_0\rangle] = N - 2 + O(1/d)$, which is close to the maximal one $M_2 = N$. Their wave function is fully extended over the many-body basis, as shown by the participation entropy $S_q(U|\Psi_0\rangle)=N + (1-q)^{-1} \log_2 \Gamma(1+q)$, i.e., the multifractal dimension is $D_q=1$. Hence, according to our definition in this work, the random Haar states are not multifractal. At the same time, the wave function of a random Haar state does not have a flat participation distribution. The multifractality flatness is non-vanishing, $\mathcal F(U|\Psi_0\rangle) >0$, and $S_q(U|\Psi_0\rangle)$ depends non-trivially on the index $q$ via the sub-leading term $c_q$ in its system size dependence, consistently with their nonstabilizerness.

\section{Probing nonstabilizerness via multifractal flatness}
\label{sec:measuring}

\subsection{Numerical example}
Here, we show that sampling over the Clifford group in Eq.~\eqref{eq:thm} can be performed via action of a circuit consisting of local Clifford gates acting on the state of interest $\ket{\Psi}$. We consider a system of $N\in[2,10]$ qubits, and, as the initial state $\ket{\Psi}$, we take either the random Haar state $U\ket{\Psi_0}$, or a product state $\ket{\Psi_1}^{\otimes N}$ of $N$ single qubit states \eqref{eq:singleQubit} with $\phi=\pi/4$ and $\theta = \pi/2$. 
We assume that the qubits form a $1$-dimensional lattice, select random site $i$, and act with a random $2$-qubit Clifford gate $U_2$~\cite{Bravyi_2021, gidney2021stimfaststabilizer} on sites $i$ and $i+1$, imposing periodic boundary conditions. 
After each action of the Clifford gate, we compute the multifractal flatness $\mathcal F$ of the obtained state. By repeating this process, we obtain results $\mathcal{F}_1,\mathcal{F}_2,\dots, \mathcal{F}_{\mathcal{N}_{\mathrm{real}} }$ which we average to calculate an estimator $\hat {\mathcal F} = \sum_{i}\mathcal F_1 /\mathcal{N}_{\mathrm{real}}$ of the mean multifractal flatness $\overline{ \mathcal{F}}(\ket{\Psi})$ over the Clifford orbit $\mathcal{C}_{\Psi}$.

Reverting Eq.~\eqref{eq:thm}, we find that 
\begin{equation}
M_2( \ket{\Psi } )= - \log_2\left[ 1- \frac{(d+1)(d+2)}{2} \hat{\mathcal{F}}\right].
\label{eq:magest}
\end{equation}
This allows us to calculate the stabilizer entropy $M_2(\ket{\Psi})$ using $\hat{\mathcal F}$, but also shows that the statistical uncertainty $\sigma(\mathcal F) $ of the multifractal flatness has to be multiplied by a factor exponential in the number of qubits $N$ to yield the statistical uncertainty $\sigma( M_2)$ of $M_2$. For instance, the results from the preceding Section for random Haar states imply that $\sigma(\mathcal F) \sim d^{5/2}/\sqrt{\mathcal N}$, which leads to $\sigma( M_2) \sim \sqrt{d/\mathcal N}$. This, in turn, indicates that in order to keep the same uncertainty of $M_2$ with increasing system size, we need to increase the number $\mathcal N$ of the samples proportionally to the Hilbert space dimension $d$. Numerical results shown in the upper panel of Fig.~\ref{fig:ex1} confirm this expectation. Keeping the same uncertainty of the stabilizer entropy with $\ket{\Psi_1}^{\otimes N}$ as the initial state requires
number of samples that is smaller but still exponential in the number of qubits: $\mathcal N \sim 2^{cN}$ (since $c\approx 0.7$).

We consider also an alternative protocol in which each calculation of the multifractal flatness $\mathcal F$ is preceded by an action of a layer of $L/2$ random 2-qubit Clifford gates. The Clifford gates from each layer act on pairs of neighboring sites covering the whole $N$ qubit chain, and the subsequent layers are shifted by $1$ site with respect to each other. Intuitively, this protocol may allow for a more efficient exploration of the Clifford orbit $\mathcal C_{\Psi}$ and a faster estimation of the stabilizer entropy in \eqref{eq:magest}. This intuition is confirmed by a comparison of the  values of $\mathcal{N}_{0.1}$ in the insets of Fig.~\ref{fig:ex1} for the two protocols. However, the asymptotic exponential scaling of the number of samples required to achieve a prescribed accuracy of $M_2( \ket{\Psi } )$  remains the same for the two protocols: $\mathcal N \sim 2^{cN}$.

Results of this numerical demonstration provide a quantitative confirmation of our analytical calculations that relate the stabilizer entropy to the multifractal flatness of the many-body wave function. Moreover, the action of local $2$-body Clifford gates is sufficient to probe the Clifford orbit $\mathcal C_{\Psi}$ and obtain an accurate estimation of the multifractal flatness $\overline{ \mathcal F}(\ket{\Psi})$. Nevertheless, the sensitivity of the stabilizer entropy to the statistical uncertainty of $\overline{ \mathcal F}(\ket{\Psi})$ makes this method of calculation of $M_2(\ket{\Psi})$ impractical -- the computational resources needed to compute $M_2(\ket{\Psi})$ for larger numbers of qubits with a given accuracy are comparable or larger than a direct calculation of $M_2(\ket{\Psi})$ according to the definition \eqref{eq:se}\footnote{We note that the resources for calculation of $M_2(\ket{\Psi})$ directly according to the definition \eqref{eq:se} scale as $O(2^{3N})$, while the cost of Monte Carlo sampling of the Clifford orbit $O(2^{2N})$ until a prescribed accuracy of $M_2(\ket{\Psi})$ is achieved scales as $O(2^{2N})$. Nevertheless, the constant describing the scaling in the Monte Carlo sampling case is large, and direct numerical calculation of $M_2(\ket{\Psi})$ according to \eqref{eq:se} is more efficient at practically relevant system sizes $N\lesssim 14$.}.

\begin{figure}[t]
    \centering
    \includegraphics[width=\columnwidth]{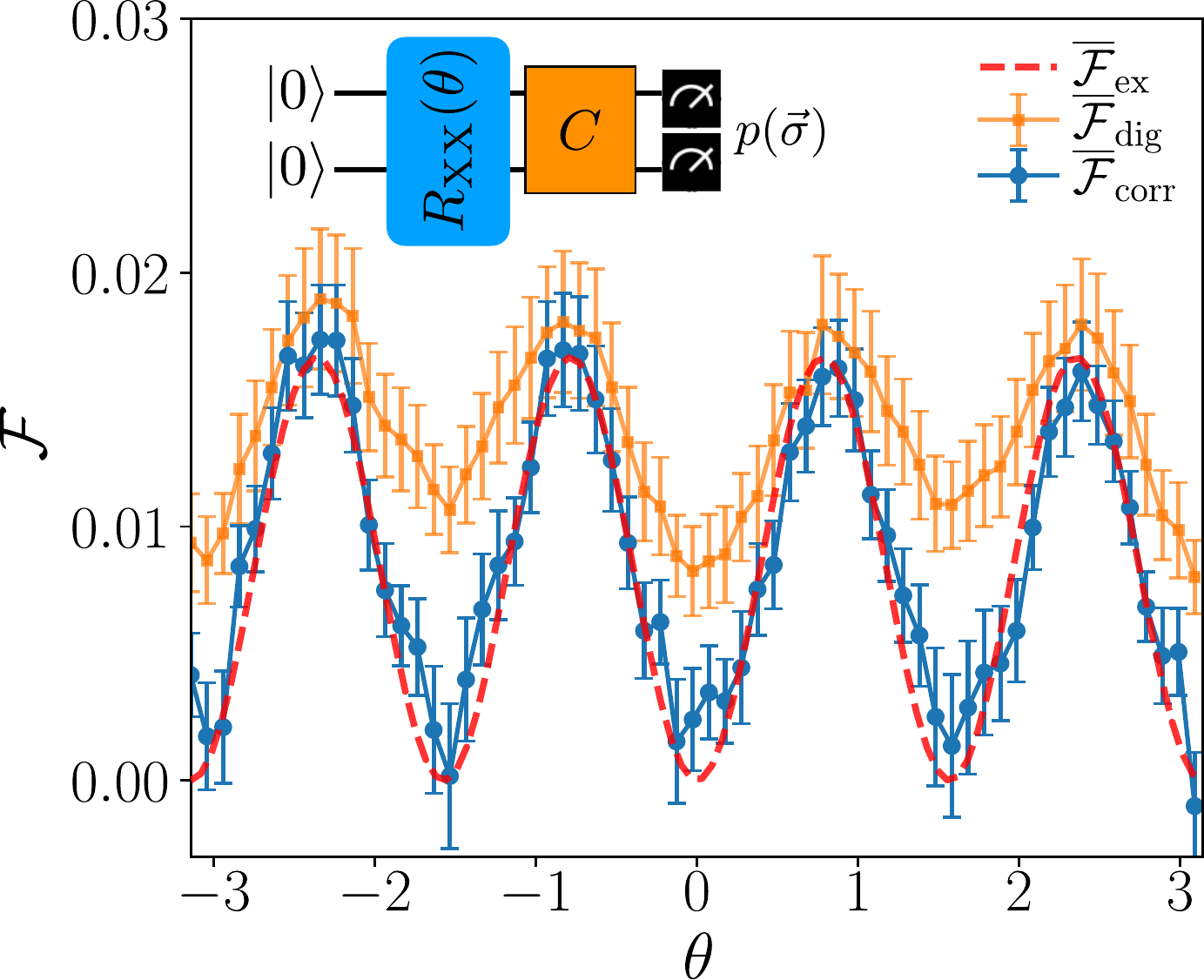}
    \caption{Digital quantum simulation of the multifractal flatness $\overline{\mathcal{F}}$ using the IBM \textsc{ibmq\_oslo} transmon quantum device. The initial state is prepared by acting with $R_\mathrm{XX}(\theta)$ gate on $|00\rangle$ state (see text). Application of a random Clifford gate $C$ followed by a read-out allows for a Monte Carlo estimate \eqref{eq:magest} of the multifractal flatness $\overline{\mathcal{F}}_\mathrm{dig}$. 
    While the features in $\theta$ are qualitatively captured, we notice a systematic shift of the quantum demonstration result $\overline{\mathcal{F}}_\mathrm{dig}$ with respect to exact analytical value $\overline{\mathcal{F}}_\mathrm{ex}$, due to imperfections of the IBM machine. Those errors can be be mitigated resulting in the corrected value of multifractal flatness $\overline{\mathcal{F}}_\mathrm{corr}$ which agrees quantitatively with $\overline{\mathcal{F}}_\mathrm{ex}$ for all values of $\theta$ considered. The error-bars show the statistical uncertainty of the results associated with number of circuit realizations which was fixed as 60. 
    }
    \label{fig:ibm}
\end{figure}

\subsection{Two-qubit system on quantum device}
Lastly, we demonstrate that the multifractal flatness $\overline{\mathcal{F}}$ is a quantity observable in current quantum devices. As a simple proof of principle, we consider a $N=2$ qubit system prepared in state $|\Psi_\theta\rangle =R_\mathrm{XX}(\theta) |\Psi_0\rangle$, where $ |\Psi_0\rangle= |00\rangle$ and $R_\mathrm{XX}(\theta) = \exp[-i \theta/2 \sigma^1\otimes \sigma^1]$. We simulate the system in the IBM transmon quantum device~\cite{ibmquantum}.
For specific angles $\theta=k \pi /2$ (where $k$ is integer) the initial state $|\Psi_\theta\rangle$ is a stabilizer, otherwise it contains nonstabilizerness. We estimate the multifractal flatness $\overline{\mathcal{F}}(|\Psi_\theta\rangle) = \sum_{C_i} \mathcal{F}(C_i|\Psi_\theta\rangle)/\mathcal{N}_\mathrm{real}$ in the \textsc{ibmq\_oslo} device using $\mathcal{N}_\mathrm{real}=60$ realizations of random Clifford gates $C_i$. 
Our results are shown in Fig.~\ref{fig:ibm}. 

The results of digital simulations, $\overline{\mathcal{F}}_\mathrm{dig}$ (denoted with orange dots), quantitatively agree with the exact analytic prediction $\overline{\mathcal{F}}_\mathrm{ex}$ in the vicinity of maxima of the multifractal flatness, where the stabilizer entropy of $|\Psi_\theta\rangle$ is close to maximal. However, in neighbourhoods of minima of $\overline{\mathcal{F}}_\mathrm{ex}$, around $\theta=k \pi /2$ (for $k$ integer), we observe a systematic deviation of  $\overline{\mathcal{F}}_\mathrm{dig}$ from the exact value. Nevertheless, we would like to emphasize that even the bare results $\overline{\mathcal{F}}_\mathrm{dig}$
qualitatively describe the variation of the multifractal flatness with the value of $\theta$.

The remaining discrepancies are due to 
a combination of decoherence and leakage errors, imperfect fidelity of gates and read-out errors. A closer inspection of the obtained vectors of quasi-probabilities $p_{\mathrm{noisy}}(\vec \sigma)$ (which approximate the probabilities $p_{\mathrm{ideal}}(\vec \sigma)=| \braket{\vec\sigma| C|\Psi_\theta\rangle } |^2$) reveals that the errors in $p_{\mathrm{noisy}}$ affect the value of the multifractal most severely for $\theta$ around $k \pi /2$. In those instances, the final state $C|\Psi_\theta\rangle$ is likely to have  $p_{\mathrm{ideal}}(\vec \sigma) =0$ for two or three basis states $\ket{ \vec \sigma}$. An error of order $\epsilon$ in $p_{\mathrm{noisy}}$ leads to an error of $\epsilon$ in the value of multifractal flatness. Hence, already errors on the level of a few percent in the quasiprobability vector $p_{\mathrm{noisy}}$ lead to error in $\mathcal F$ on the level of the signal in Fig.~\ref{fig:ibm}.
 In contrast,  around $\theta=(2k+1) \pi /4$, the probability $p_{\mathrm{ideal}}(\vec \sigma)$ is distributed more uniformly over the basis $\ket{ \vec \sigma}$ for majority of the Clifford gates $C$. Hence, some of the errors in $p_{\mathrm{noisy}}$ propagate only at the second order in $\epsilon$, while some errors cancel out, yielding $\overline{\mathcal{F}}_\mathrm{dig}$  close to the exact value.


Below, we show that a passive readout error mitigation scheme \cite{Geller20} is sufficient to correct the bare results $\overline{\mathcal{F}}_\mathrm{dig}$.
We fix the Clifford gate $C$ to be equal to unity, and set $\theta=0.01$. Considering all $4=2^N$ ($N=2$) initial states, we obtain the quasiprobability vectors $p_{\mathrm{noisy}}$ and find matrix $A$ which links them with the ideal results  $p_{\mathrm{ideal}}$:
\begin{equation}
p_{\mathrm{noisy}}(\vec \sigma) = A \, p_{\mathrm{ideal}}(\vec \sigma).
\label{eq:pnoi}
\end{equation}
We observe that the $4\times 4 $ matrix $A$ can be parametrizedin the basis $\{ \ket{\vec \sigma} \}= \{ \ket{01},\ket{00},\ket{10},\ket{11}\}$ with good accuracy as 
\begin{equation}
 A=  
  \begin{bmatrix}
    1-2p-q & p & p & q \\
    p & 1-2p-q & q & p \\
    p & q & 1-2p-q & p \\
    q & p & p & 1-2p-q
  \end{bmatrix},
  \label{eq:amat}
\end{equation}
where $p$ and $q$ are parameters characteristic for a given device. For \textsc{ibmq\_oslo} we find that the results are broadly consistent with $p=0.045$ and $q=0.02$. Inverting the formula \eqref{eq:pnoi}, given the matrix $A$, allows us to find an estimate for $p_{\mathrm{ideal}}(\vec \sigma)$. Performing this error-mitigation for the data gathered for arbitrary $\theta$ and $C$, we obtain the corrected value of multifractal flatness $\overline{\mathcal{F}}_\mathrm{corr}$, which agrees, within the error-bars associated with sampling over the Clifford orbit, with the exact result  $\overline{\mathcal{F}}_\mathrm{ex}$. 

The presented results illustrate that the multifractal flatness $\overline{\mathcal{F}}$ is indeed measurable on current quantum devices, albeit for very small number of qubit $N=2$. Finding more controlled and scalable error-mitigation schemes is an interesting future challange, which might involve use of the 
zero noise extrapolation \cite{Giurgica20} and active error mitigation techniques \cite{Nation21, Alistair21, Hicks22}.


\section{Conclusion}
This work puts forward a correspondence between the concept of wave function multifractality and nonstabilizerness. Specifically, we introduce multifractal flatness, a combination of inverse participation ratios that quantifies the flatness of the wave function probability distribution. We show that multifractal flatness is a witness of nonstabilizerness and translates to a measure of magic when the average over the Clifford orbit is considered.  
We illustrate the relationship between multifractality and nonstabilizerness on a few examples of quantum states, for instance recovering known results~\cite{leone2022stabilizerrenyientropy} about random Haar states. 

The connection put forward in this work has practical implications, as we show by computing the stabilizer entropies by estimating the multifractal flatness with a repeated action of a local Clifford circuit.
As a proof of principle, we demonstrate that nonstabilizerness, as estimated by the multifractal flatness $\overline{\mathcal{F}}$, is detectable in current quantum devices. 

Recent works~\cite{leone2021quantumchaosis,leone2022learning} revealed a tight relationship between quantum chaos and nonstabilizerness. It would be interesting to investigate this link through the lens of multifractality. For instance, in semiclassical quantum systems, the generalized Lyapunov exponents present multifractal features~\cite{Pappalardi2023,pap3}. Furthermore, models of maximally chaotic quantum many-body systems have been shown to display rich multifractal behavior~\cite{micklitz2019nonergodic,monteiro2021minimal,dieplinger2021ansykinspired}. Additionally, it would be interesting to reveal the equilibrium and out-of-equilibrium nonstabilizerness of archetypal condensed matter models~\cite{leone2023nonstabilizerness,lami2023quantum,oliviero2021transitionsinentanglementcomplexity}.

\textit{Note added: Recently, a manuscript appeared~\cite{haug2023efficient} implementing a scalable scheme for experimentally observing the multifractal flatness.
}

\textit{Acknowledgments.---} We thank L. Piroli,  T. Chanda, M. Dalmonte, G. Fux, and R. Fazio for enlightening discussions. X.T. is indebted to S. Pappalardi for consultations on magic, multifractal, and to G. M. Andolina for discussions.
We acknowledge the use of IBM Quantum services for this work and to advanced services provided by the IBM Quantum Researchers Program. The views expressed are those of the authors, and do not reflect the official policy or position of IBM or the IBM Quantum team.
We acknowledge the workshop "Dynamical Foundation of Many-Body Quantum Chaos" at Institute Pascal (Orsay, France) for hosting us while finalizing this manuscript's writing.
X.T. and M. S. acknowledge support from the ANR grant “NonEQuMat” (ANR-19-CE47-0001). 
P.S. acknowledges support from: ERC AdG NOQIA; Ministerio de Ciencia y Innovation Agencia Estatal de Investigaciones (PGC2018-097027-B-I00/10.13039/501100011033, CEX2019-000910-S/10.13039/501100011033, Plan National FIDEUA PID2019-106901GB-I00, FPI, QUANTERA MAQS PCI2019-111828-2, QUANTERA DYNAMITE PCI2022-132919, Proyectos de I+D+I “Retos Colaboración” QUSPIN RTC2019-007196-7); MICIIN with funding from European Union NextGenerationEU(PRTR-C17.I1) and by Generalitat de Catalunya; Fundació Cellex; Fundació Mir-Puig; Generalitat de Catalunya (European Social Fund FEDER and CERCA program, AGAUR Grant No. 2021 SGR 01452, QuantumCAT \ U16-011424, co-funded by ERDF Operational Program of Catalonia 2014-2020); Barcelona Supercomputing Center MareNostrum (FI-2023-1-0013); EU Horizon 2020 FET-OPEN OPTOlogic (Grant No 899794); EU Horizon Europe Program (Grant Agreement 101080086 — NeQST), National Science Centre, Poland (Symfonia Grant No. 2016/20/W/ST4/00314); ICFO Internal “QuantumGaudi” project; European Union’s Horizon 2020 research and innovation program under the Marie-Skłodowska-Curie grant agreement No 101029393 (STREDCH) and No 847648 (“La Caixa” Junior Leaders fellowships ID100010434: LCF/BQ/PI19/11690013, LCF/BQ/PI20/11760031, LCF/BQ/PR20/11770012, LCF/BQ/PR21/11840013). 
Views and opinions expressed in this work are, however, those of the author(s) only and do not necessarily reflect those of the European Union, European Climate, Infrastructure and Environment Executive Agency (CINEA), nor any other granting authority. Neither the European Union nor any granting authority can be held responsible for them.


\begin{thebibliography}{110}%
\makeatletter
\providecommand \@ifxundefined [1]{%
 \@ifx{#1\undefined}
}%
\providecommand \@ifnum [1]{%
 \ifnum #1\expandafter \@firstoftwo
 \else \expandafter \@secondoftwo
 \fi
}%
\providecommand \@ifx [1]{%
 \ifx #1\expandafter \@firstoftwo
 \else \expandafter \@secondoftwo
 \fi
}%
\providecommand \natexlab [1]{#1}%
\providecommand \enquote  [1]{``#1''}%
\providecommand \bibnamefont  [1]{#1}%
\providecommand \bibfnamefont [1]{#1}%
\providecommand \citenamefont [1]{#1}%
\providecommand \href@noop [0]{\@secondoftwo}%
\providecommand \href [0]{\begingroup \@sanitize@url \@href}%
\providecommand \@href[1]{\@@startlink{#1}\@@href}%
\providecommand \@@href[1]{\endgroup#1\@@endlink}%
\providecommand \@sanitize@url [0]{\catcode `\\12\catcode `\$12\catcode `\&12\catcode `\#12\catcode `\^12\catcode `\_12\catcode `\%12\relax}%
\providecommand \@@startlink[1]{}%
\providecommand \@@endlink[0]{}%
\providecommand \url  [0]{\begingroup\@sanitize@url \@url }%
\providecommand \@url [1]{\endgroup\@href {#1}{\urlprefix }}%
\providecommand \urlprefix  [0]{URL }%
\providecommand \Eprint [0]{\href }%
\providecommand \doibase [0]{https://doi.org/}%
\providecommand \selectlanguage [0]{\@gobble}%
\providecommand \bibinfo  [0]{\@secondoftwo}%
\providecommand \bibfield  [0]{\@secondoftwo}%
\providecommand \translation [1]{[#1]}%
\providecommand \BibitemOpen [0]{}%
\providecommand \bibitemStop [0]{}%
\providecommand \bibitemNoStop [0]{.\EOS\space}%
\providecommand \EOS [0]{\spacefactor3000\relax}%
\providecommand \BibitemShut  [1]{\csname bibitem#1\endcsname}%
\let\auto@bib@innerbib\@empty
\bibitem [{\citenamefont {Gottesman}(1998)}]{gottesman1998theoryoffaulttolerant}%
  \BibitemOpen
  \bibfield  {author} {\bibinfo {author} {\bibfnamefont {D.}~\bibnamefont {Gottesman}},\ }\href {https://doi.org/10.1103/PhysRevA.57.127} {\bibfield  {journal} {\bibinfo  {journal} {Phys. Rev. A}\ }\textbf {\bibinfo {volume} {57}},\ \bibinfo {pages} {127} (\bibinfo {year} {1998})}\BibitemShut {NoStop}%
\bibitem [{\citenamefont {Eastin}\ and\ \citenamefont {Knill}(2009)}]{eastin2009restrictions}%
  \BibitemOpen
  \bibfield  {author} {\bibinfo {author} {\bibfnamefont {B.}~\bibnamefont {Eastin}}\ and\ \bibinfo {author} {\bibfnamefont {E.}~\bibnamefont {Knill}},\ }\href {https://doi.org/10.1103/PhysRevLett.102.110502} {\bibfield  {journal} {\bibinfo  {journal} {Phys. Rev. Lett.}\ }\textbf {\bibinfo {volume} {102}},\ \bibinfo {pages} {110502} (\bibinfo {year} {2009})}\BibitemShut {NoStop}%
\bibitem [{\citenamefont {Nielsen}\ and\ \citenamefont {Chuang}(2000)}]{nielsen00}%
  \BibitemOpen
  \bibfield  {author} {\bibinfo {author} {\bibfnamefont {M.~A.}\ \bibnamefont {Nielsen}}\ and\ \bibinfo {author} {\bibfnamefont {I.~L.}\ \bibnamefont {Chuang}},\ }\href@noop {} {\emph {\bibinfo {title} {Quantum Computation and Quantum Information}}}\ (\bibinfo  {publisher} {Cambridge University Press},\ \bibinfo {year} {2000})\BibitemShut {NoStop}%
\bibitem [{\citenamefont {Fritzsch}\ and\ \citenamefont {Prosen}(2021)}]{Fritzsch21}%
  \BibitemOpen
  \bibfield  {author} {\bibinfo {author} {\bibfnamefont {F.}~\bibnamefont {Fritzsch}}\ and\ \bibinfo {author} {\bibfnamefont {T.}~\bibnamefont {Prosen}},\ }\href {https://doi.org/10.1103/PhysRevE.103.062133} {\bibfield  {journal} {\bibinfo  {journal} {Phys. Rev. E}\ }\textbf {\bibinfo {volume} {103}},\ \bibinfo {pages} {062133} (\bibinfo {year} {2021})}\BibitemShut {NoStop}%
\bibitem [{\citenamefont {Claeys}\ and\ \citenamefont {Lamacraft}(2021)}]{Claeys21}%
  \BibitemOpen
  \bibfield  {author} {\bibinfo {author} {\bibfnamefont {P.~W.}\ \bibnamefont {Claeys}}\ and\ \bibinfo {author} {\bibfnamefont {A.}~\bibnamefont {Lamacraft}},\ }\href {https://doi.org/10.1103/PhysRevLett.126.100603} {\bibfield  {journal} {\bibinfo  {journal} {Phys. Rev. Lett.}\ }\textbf {\bibinfo {volume} {126}},\ \bibinfo {pages} {100603} (\bibinfo {year} {2021})}\BibitemShut {NoStop}%
\bibitem [{\citenamefont {Nahum}\ \emph {et~al.}(2017)\citenamefont {Nahum}, \citenamefont {Ruhman}, \citenamefont {Vijay},\ and\ \citenamefont {Haah}}]{nahum2017quantum}%
  \BibitemOpen
  \bibfield  {author} {\bibinfo {author} {\bibfnamefont {A.}~\bibnamefont {Nahum}}, \bibinfo {author} {\bibfnamefont {J.}~\bibnamefont {Ruhman}}, \bibinfo {author} {\bibfnamefont {S.}~\bibnamefont {Vijay}},\ and\ \bibinfo {author} {\bibfnamefont {J.}~\bibnamefont {Haah}},\ }\href {https://doi.org/10.1103/PhysRevX.7.031016} {\bibfield  {journal} {\bibinfo  {journal} {Phys. Rev. X}\ }\textbf {\bibinfo {volume} {7}},\ \bibinfo {pages} {031016} (\bibinfo {year} {2017})}\BibitemShut {NoStop}%
\bibitem [{\citenamefont {Nahum}\ \emph {et~al.}(2018)\citenamefont {Nahum}, \citenamefont {Vijay},\ and\ \citenamefont {Haah}}]{nahum2018operator}%
  \BibitemOpen
  \bibfield  {author} {\bibinfo {author} {\bibfnamefont {A.}~\bibnamefont {Nahum}}, \bibinfo {author} {\bibfnamefont {S.}~\bibnamefont {Vijay}},\ and\ \bibinfo {author} {\bibfnamefont {J.}~\bibnamefont {Haah}},\ }\href {https://doi.org/10.1103/PhysRevX.8.021014} {\bibfield  {journal} {\bibinfo  {journal} {Phys. Rev. X}\ }\textbf {\bibinfo {volume} {8}},\ \bibinfo {pages} {021014} (\bibinfo {year} {2018})}\BibitemShut {NoStop}%
\bibitem [{\citenamefont {Fisher}\ \emph {et~al.}(2023)\citenamefont {Fisher}, \citenamefont {Khemani}, \citenamefont {Nahum},\ and\ \citenamefont {Vijay}}]{Fisher2023}%
  \BibitemOpen
  \bibfield  {author} {\bibinfo {author} {\bibfnamefont {M.~P.}\ \bibnamefont {Fisher}}, \bibinfo {author} {\bibfnamefont {V.}~\bibnamefont {Khemani}}, \bibinfo {author} {\bibfnamefont {A.}~\bibnamefont {Nahum}},\ and\ \bibinfo {author} {\bibfnamefont {S.}~\bibnamefont {Vijay}},\ }\href {https://doi.org/10.1146/annurev-conmatphys-031720-030658} {\bibfield  {journal} {\bibinfo  {journal} {Annu. Rev. Condens. Matter Phys.}\ }\textbf {\bibinfo {volume} {14}},\ \bibinfo {pages} {335} (\bibinfo {year} {2023})}\BibitemShut {NoStop}%
\bibitem [{\citenamefont {Pastawski}\ \emph {et~al.}(2015)\citenamefont {Pastawski}, \citenamefont {Yoshida}, \citenamefont {Harlow},\ and\ \citenamefont {Preskill}}]{Pastawski2015}%
  \BibitemOpen
  \bibfield  {author} {\bibinfo {author} {\bibfnamefont {F.}~\bibnamefont {Pastawski}}, \bibinfo {author} {\bibfnamefont {B.}~\bibnamefont {Yoshida}}, \bibinfo {author} {\bibfnamefont {D.}~\bibnamefont {Harlow}},\ and\ \bibinfo {author} {\bibfnamefont {J.}~\bibnamefont {Preskill}},\ }\href {https://doi.org/10.1007/jhep06(2015)149} {\bibfield  {journal} {\bibinfo  {journal} {J. High Energy Phys.}\ }\textbf {\bibinfo {volume} {2015}}\bibinfo  {number} { (6)}}\BibitemShut {NoStop}%
\bibitem [{\citenamefont {Aaronson}\ and\ \citenamefont {Gottesman}(2004)}]{aaronson2004improvedsimulationof}%
  \BibitemOpen
\bibfield  {number} {  }\bibfield  {author} {\bibinfo {author} {\bibfnamefont {S.}~\bibnamefont {Aaronson}}\ and\ \bibinfo {author} {\bibfnamefont {D.}~\bibnamefont {Gottesman}},\ }\href {https://doi.org/10.1103/PhysRevA.70.052328} {\bibfield  {journal} {\bibinfo  {journal} {Phys. Rev. A}\ }\textbf {\bibinfo {volume} {70}},\ \bibinfo {pages} {052328} (\bibinfo {year} {2004})}\BibitemShut {NoStop}%
\bibitem [{\citenamefont {Shor}(1997)}]{shor1997polynomialtimealgorithmsfor}%
  \BibitemOpen
  \bibfield  {author} {\bibinfo {author} {\bibfnamefont {P.~W.}\ \bibnamefont {Shor}},\ }\href {https://doi.org/10.1137/s0097539795293172} {\bibfield  {journal} {\bibinfo  {journal} {{SIAM} Journal on Computing}\ }\textbf {\bibinfo {volume} {26}},\ \bibinfo {pages} {1484} (\bibinfo {year} {1997})}\BibitemShut {NoStop}%
\bibitem [{\citenamefont {Feynman}(1982)}]{feynman1982simulatingphysicswith}%
  \BibitemOpen
  \bibfield  {author} {\bibinfo {author} {\bibfnamefont {R.~P.}\ \bibnamefont {Feynman}},\ }\href {https://doi.org/10.1007/bf02650179} {\bibfield  {journal} {\bibinfo  {journal} {Int. J. Theor. Phys.}\ }\textbf {\bibinfo {volume} {21}},\ \bibinfo {pages} {467} (\bibinfo {year} {1982})}\BibitemShut {NoStop}%
\bibitem [{\citenamefont {Deutsch}(1985)}]{deutsch1985quantumtheory}%
  \BibitemOpen
  \bibfield  {author} {\bibinfo {author} {\bibfnamefont {D.}~\bibnamefont {Deutsch}},\ }\href {https://doi.org/10.1098/rspa.1985.0070} {\bibfield  {journal} {\bibinfo  {journal} {Proc. R. Soc. A: Math. Phys. Eng. Sci.}\ }\textbf {\bibinfo {volume} {400}},\ \bibinfo {pages} {97} (\bibinfo {year} {1985})}\BibitemShut {NoStop}%
\bibitem [{\citenamefont {Kitaev}(2003)}]{Kitaev2003}%
  \BibitemOpen
  \bibfield  {author} {\bibinfo {author} {\bibfnamefont {A.}~\bibnamefont {Kitaev}},\ }\href {https://doi.org/10.1016/s0003-4916(02)00018-0} {\bibfield  {journal} {\bibinfo  {journal} {Ann. Phys.}\ }\textbf {\bibinfo {volume} {303}},\ \bibinfo {pages} {2} (\bibinfo {year} {2003})}\BibitemShut {NoStop}%
\bibitem [{\citenamefont {Gottesman}\ and\ \citenamefont {Chuang}(1999)}]{Gottesman1999demonstratingtheviability}%
  \BibitemOpen
  \bibfield  {author} {\bibinfo {author} {\bibfnamefont {D.}~\bibnamefont {Gottesman}}\ and\ \bibinfo {author} {\bibfnamefont {I.~L.}\ \bibnamefont {Chuang}},\ }\href {https://doi.org/10.1038/46503} {\bibfield  {journal} {\bibinfo  {journal} {Nature}\ }\textbf {\bibinfo {volume} {402}},\ \bibinfo {pages} {390} (\bibinfo {year} {1999})}\BibitemShut {NoStop}%
\bibitem [{\citenamefont {Bravyi}\ and\ \citenamefont {Kitaev}(2005)}]{bravyi2005universalquantumcomputation}%
  \BibitemOpen
  \bibfield  {author} {\bibinfo {author} {\bibfnamefont {S.}~\bibnamefont {Bravyi}}\ and\ \bibinfo {author} {\bibfnamefont {A.}~\bibnamefont {Kitaev}},\ }\href {https://doi.org/10.1103/PhysRevA.71.022316} {\bibfield  {journal} {\bibinfo  {journal} {Phys. Rev. A}\ }\textbf {\bibinfo {volume} {71}},\ \bibinfo {pages} {022316} (\bibinfo {year} {2005})}\BibitemShut {NoStop}%
\bibitem [{\citenamefont {Veitch}\ \emph {et~al.}(2014)\citenamefont {Veitch}, \citenamefont {Mousavian}, \citenamefont {Gottesman},\ and\ \citenamefont {Emerson}}]{Veitch2014theresourcetheory}%
  \BibitemOpen
  \bibfield  {author} {\bibinfo {author} {\bibfnamefont {V.}~\bibnamefont {Veitch}}, \bibinfo {author} {\bibfnamefont {S.~A.~H.}\ \bibnamefont {Mousavian}}, \bibinfo {author} {\bibfnamefont {D.}~\bibnamefont {Gottesman}},\ and\ \bibinfo {author} {\bibfnamefont {J.}~\bibnamefont {Emerson}},\ }\href {https://doi.org/10.1088/1367-2630/16/1/013009} {\bibfield  {journal} {\bibinfo  {journal} {New J. Phys.}\ }\textbf {\bibinfo {volume} {16}},\ \bibinfo {pages} {013009} (\bibinfo {year} {2014})}\BibitemShut {NoStop}%
\bibitem [{\citenamefont {Bravyi}\ \emph {et~al.}(2016)\citenamefont {Bravyi}, \citenamefont {Smith},\ and\ \citenamefont {Smolin}}]{bravyi2016tradingclassicaland}%
  \BibitemOpen
  \bibfield  {author} {\bibinfo {author} {\bibfnamefont {S.}~\bibnamefont {Bravyi}}, \bibinfo {author} {\bibfnamefont {G.}~\bibnamefont {Smith}},\ and\ \bibinfo {author} {\bibfnamefont {J.~A.}\ \bibnamefont {Smolin}},\ }\href {https://doi.org/10.1103/PhysRevX.6.021043} {\bibfield  {journal} {\bibinfo  {journal} {Phys. Rev. X}\ }\textbf {\bibinfo {volume} {6}},\ \bibinfo {pages} {021043} (\bibinfo {year} {2016})}\BibitemShut {NoStop}%
\bibitem [{\citenamefont {Chitambar}\ and\ \citenamefont {Gour}(2019)}]{chitambar2019quantumresourcetheories}%
  \BibitemOpen
  \bibfield  {author} {\bibinfo {author} {\bibfnamefont {E.}~\bibnamefont {Chitambar}}\ and\ \bibinfo {author} {\bibfnamefont {G.}~\bibnamefont {Gour}},\ }\href {https://doi.org/10.1103/RevModPhys.91.025001} {\bibfield  {journal} {\bibinfo  {journal} {Rev. Mod. Phys.}\ }\textbf {\bibinfo {volume} {91}},\ \bibinfo {pages} {025001} (\bibinfo {year} {2019})}\BibitemShut {NoStop}%
\bibitem [{\citenamefont {Kelly}\ \emph {et~al.}()\citenamefont {Kelly}, \citenamefont {Poschinger}, \citenamefont {Schmidt-Kaler}, \citenamefont {Fisher},\ and\ \citenamefont {Marino}}]{kelly2023coherence}%
  \BibitemOpen
  \bibfield  {author} {\bibinfo {author} {\bibfnamefont {S.~P.}\ \bibnamefont {Kelly}}, \bibinfo {author} {\bibfnamefont {U.}~\bibnamefont {Poschinger}}, \bibinfo {author} {\bibfnamefont {F.}~\bibnamefont {Schmidt-Kaler}}, \bibinfo {author} {\bibfnamefont {M.~P.~A.}\ \bibnamefont {Fisher}},\ and\ \bibinfo {author} {\bibfnamefont {J.}~\bibnamefont {Marino}},\ }\href@noop {} {}\Eprint {https://arxiv.org/abs/2210.11547} {arXiv:2210.11547} \BibitemShut {NoStop}%
\bibitem [{\citenamefont {Weinstein}\ \emph {et~al.}()\citenamefont {Weinstein}, \citenamefont {Kelly}, \citenamefont {Marino},\ and\ \citenamefont {Altman}}]{weinstein2022scrambling}%
  \BibitemOpen
  \bibfield  {author} {\bibinfo {author} {\bibfnamefont {Z.}~\bibnamefont {Weinstein}}, \bibinfo {author} {\bibfnamefont {S.~P.}\ \bibnamefont {Kelly}}, \bibinfo {author} {\bibfnamefont {J.}~\bibnamefont {Marino}},\ and\ \bibinfo {author} {\bibfnamefont {E.}~\bibnamefont {Altman}},\ }\href@noop {} {}\Eprint {https://arxiv.org/abs/2210.14242} {arXiv:2210.14242} \BibitemShut {NoStop}%
\bibitem [{\citenamefont {Howard}\ \emph {et~al.}(2014)\citenamefont {Howard}, \citenamefont {Wallman}, \citenamefont {Veitch},\ and\ \citenamefont {Emerson}}]{Howard2014}%
  \BibitemOpen
  \bibfield  {author} {\bibinfo {author} {\bibfnamefont {M.}~\bibnamefont {Howard}}, \bibinfo {author} {\bibfnamefont {J.}~\bibnamefont {Wallman}}, \bibinfo {author} {\bibfnamefont {V.}~\bibnamefont {Veitch}},\ and\ \bibinfo {author} {\bibfnamefont {J.}~\bibnamefont {Emerson}},\ }\href {https://doi.org/10.1038/nature13460} {\bibfield  {journal} {\bibinfo  {journal} {Nature}\ }\textbf {\bibinfo {volume} {510}},\ \bibinfo {pages} {351} (\bibinfo {year} {2014})}\BibitemShut {NoStop}%
\bibitem [{\citenamefont {Seddon}\ \emph {et~al.}(2021)\citenamefont {Seddon}, \citenamefont {Regula}, \citenamefont {Pashayan}, \citenamefont {Ouyang},\ and\ \citenamefont {Campbell}}]{seddon2021quantifying}%
  \BibitemOpen
  \bibfield  {author} {\bibinfo {author} {\bibfnamefont {J.~R.}\ \bibnamefont {Seddon}}, \bibinfo {author} {\bibfnamefont {B.}~\bibnamefont {Regula}}, \bibinfo {author} {\bibfnamefont {H.}~\bibnamefont {Pashayan}}, \bibinfo {author} {\bibfnamefont {Y.}~\bibnamefont {Ouyang}},\ and\ \bibinfo {author} {\bibfnamefont {E.~T.}\ \bibnamefont {Campbell}},\ }\href {https://doi.org/10.1103/PRXQuantum.2.010345} {\bibfield  {journal} {\bibinfo  {journal} {PRX Quantum}\ }\textbf {\bibinfo {volume} {2}},\ \bibinfo {pages} {010345} (\bibinfo {year} {2021})}\BibitemShut {NoStop}%
\bibitem [{\citenamefont {Koukoulekidis}\ \emph {et~al.}(2022)\citenamefont {Koukoulekidis}, \citenamefont {Kwon}, \citenamefont {Jee}, \citenamefont {Jennings},\ and\ \citenamefont {Kim}}]{Koukoulekidis2022fasterborn}%
  \BibitemOpen
  \bibfield  {author} {\bibinfo {author} {\bibfnamefont {N.}~\bibnamefont {Koukoulekidis}}, \bibinfo {author} {\bibfnamefont {H.}~\bibnamefont {Kwon}}, \bibinfo {author} {\bibfnamefont {H.~H.}\ \bibnamefont {Jee}}, \bibinfo {author} {\bibfnamefont {D.}~\bibnamefont {Jennings}},\ and\ \bibinfo {author} {\bibfnamefont {M.~S.}\ \bibnamefont {Kim}},\ }\href {https://doi.org/10.22331/q-2022-10-13-838} {\bibfield  {journal} {\bibinfo  {journal} {{Quantum}}\ }\textbf {\bibinfo {volume} {6}},\ \bibinfo {pages} {838} (\bibinfo {year} {2022})}\BibitemShut {NoStop}%
\bibitem [{\citenamefont {Haferkamp}(2022)}]{haferkamp2022randomquantum}%
  \BibitemOpen
  \bibfield  {author} {\bibinfo {author} {\bibfnamefont {J.}~\bibnamefont {Haferkamp}},\ }\href {https://doi.org/10.22331/q-2022-09-08-795} {\bibfield  {journal} {\bibinfo  {journal} {{Quantum}}\ }\textbf {\bibinfo {volume} {6}},\ \bibinfo {pages} {795} (\bibinfo {year} {2022})}\BibitemShut {NoStop}%
\bibitem [{\citenamefont {Leone}\ \emph {et~al.}(2021)\citenamefont {Leone}, \citenamefont {Oliviero}, \citenamefont {Zhou},\ and\ \citenamefont {Hamma}}]{leone2021quantumchaosis}%
  \BibitemOpen
  \bibfield  {author} {\bibinfo {author} {\bibfnamefont {L.}~\bibnamefont {Leone}}, \bibinfo {author} {\bibfnamefont {S.~F.~E.}\ \bibnamefont {Oliviero}}, \bibinfo {author} {\bibfnamefont {Y.}~\bibnamefont {Zhou}},\ and\ \bibinfo {author} {\bibfnamefont {A.}~\bibnamefont {Hamma}},\ }\href {https://doi.org/10.22331/q-2021-05-04-453} {\bibfield  {journal} {\bibinfo  {journal} {{Quantum}}\ }\textbf {\bibinfo {volume} {5}},\ \bibinfo {pages} {453} (\bibinfo {year} {2021})}\BibitemShut {NoStop}%
\bibitem [{\citenamefont {Campbell}(2011)}]{campbell2011catalysisandactivation}%
  \BibitemOpen
  \bibfield  {author} {\bibinfo {author} {\bibfnamefont {E.~T.}\ \bibnamefont {Campbell}},\ }\href {https://doi.org/10.1103/PhysRevA.83.032317} {\bibfield  {journal} {\bibinfo  {journal} {Phys. Rev. A}\ }\textbf {\bibinfo {volume} {83}},\ \bibinfo {pages} {032317} (\bibinfo {year} {2011})}\BibitemShut {NoStop}%
\bibitem [{\citenamefont {Howard}\ and\ \citenamefont {Campbell}(2017)}]{howard2017application}%
  \BibitemOpen
  \bibfield  {author} {\bibinfo {author} {\bibfnamefont {M.}~\bibnamefont {Howard}}\ and\ \bibinfo {author} {\bibfnamefont {E.}~\bibnamefont {Campbell}},\ }\href {https://doi.org/10.1103/PhysRevLett.118.090501} {\bibfield  {journal} {\bibinfo  {journal} {Phys. Rev. Lett.}\ }\textbf {\bibinfo {volume} {118}},\ \bibinfo {pages} {090501} (\bibinfo {year} {2017})}\BibitemShut {NoStop}%
\bibitem [{\citenamefont {Wang}\ \emph {et~al.}(2019)\citenamefont {Wang}, \citenamefont {Wilde},\ and\ \citenamefont {Su}}]{Wang2019}%
  \BibitemOpen
  \bibfield  {author} {\bibinfo {author} {\bibfnamefont {X.}~\bibnamefont {Wang}}, \bibinfo {author} {\bibfnamefont {M.~M.}\ \bibnamefont {Wilde}},\ and\ \bibinfo {author} {\bibfnamefont {Y.}~\bibnamefont {Su}},\ }\href {https://doi.org/10.1088/1367-2630/ab451d} {\bibfield  {journal} {\bibinfo  {journal} {New J. Phys.}\ }\textbf {\bibinfo {volume} {21}},\ \bibinfo {pages} {103002} (\bibinfo {year} {2019})}\BibitemShut {NoStop}%
\bibitem [{\citenamefont {Beverland}\ \emph {et~al.}(2020)\citenamefont {Beverland}, \citenamefont {Campbell}, \citenamefont {Howard},\ and\ \citenamefont {Kliuchnikov}}]{Beverland2020}%
  \BibitemOpen
  \bibfield  {author} {\bibinfo {author} {\bibfnamefont {M.}~\bibnamefont {Beverland}}, \bibinfo {author} {\bibfnamefont {E.}~\bibnamefont {Campbell}}, \bibinfo {author} {\bibfnamefont {M.}~\bibnamefont {Howard}},\ and\ \bibinfo {author} {\bibfnamefont {V.}~\bibnamefont {Kliuchnikov}},\ }\href {https://doi.org/10.1088/2058-9565/ab8963} {\bibfield  {journal} {\bibinfo  {journal} {Quantum Sci. Technol.}\ }\textbf {\bibinfo {volume} {5}},\ \bibinfo {pages} {035009} (\bibinfo {year} {2020})}\BibitemShut {NoStop}%
\bibitem [{\citenamefont {Liu}\ and\ \citenamefont {Winter}(2022)}]{liu2022manybodyquantummagic}%
  \BibitemOpen
  \bibfield  {author} {\bibinfo {author} {\bibfnamefont {Z.-W.}\ \bibnamefont {Liu}}\ and\ \bibinfo {author} {\bibfnamefont {A.}~\bibnamefont {Winter}},\ }\href {https://doi.org/10.1103/PRXQuantum.3.020333} {\bibfield  {journal} {\bibinfo  {journal} {PRX Quantum}\ }\textbf {\bibinfo {volume} {3}},\ \bibinfo {pages} {020333} (\bibinfo {year} {2022})}\BibitemShut {NoStop}%
\bibitem [{\citenamefont {Bu}\ \emph {et~al.}(2022)\citenamefont {Bu}, \citenamefont {Garcia}, \citenamefont {Jaffe}, \citenamefont {Koh},\ and\ \citenamefont {Li}}]{bu2022complexity}%
  \BibitemOpen
  \bibfield  {author} {\bibinfo {author} {\bibfnamefont {K.}~\bibnamefont {Bu}}, \bibinfo {author} {\bibfnamefont {R.~J.}\ \bibnamefont {Garcia}}, \bibinfo {author} {\bibfnamefont {A.}~\bibnamefont {Jaffe}}, \bibinfo {author} {\bibfnamefont {D.~E.}\ \bibnamefont {Koh}},\ and\ \bibinfo {author} {\bibfnamefont {L.}~\bibnamefont {Li}},\ }\href@noop {} {} (\bibinfo {year} {2022}),\ \Eprint {https://arxiv.org/abs/2204.12051} {2204.12051} \BibitemShut {NoStop}%
\bibitem [{\citenamefont {Jiang}\ and\ \citenamefont {Wang}(2023)}]{jiang2023lowerboundfor}%
  \BibitemOpen
  \bibfield  {author} {\bibinfo {author} {\bibfnamefont {J.}~\bibnamefont {Jiang}}\ and\ \bibinfo {author} {\bibfnamefont {X.}~\bibnamefont {Wang}},\ }\href {https://doi.org/10.1103/PhysRevApplied.19.034052} {\bibfield  {journal} {\bibinfo  {journal} {Phys. Rev. Appl.}\ }\textbf {\bibinfo {volume} {19}},\ \bibinfo {pages} {034052} (\bibinfo {year} {2023})}\BibitemShut {NoStop}%
\bibitem [{\citenamefont {Haug}\ and\ \citenamefont {Piroli}()}]{haug2023stabilizerentropiesand}%
  \BibitemOpen
  \bibfield  {author} {\bibinfo {author} {\bibfnamefont {T.}~\bibnamefont {Haug}}\ and\ \bibinfo {author} {\bibfnamefont {L.}~\bibnamefont {Piroli}},\ }\href@noop {} {}\Eprint {https://arxiv.org/abs/2303.10152} {arXiv:2303.10152} \BibitemShut {NoStop}%
\bibitem [{\citenamefont {Leone}\ \emph {et~al.}(2022)\citenamefont {Leone}, \citenamefont {Oliviero},\ and\ \citenamefont {Hamma}}]{leone2022stabilizerrenyientropy}%
  \BibitemOpen
  \bibfield  {author} {\bibinfo {author} {\bibfnamefont {L.}~\bibnamefont {Leone}}, \bibinfo {author} {\bibfnamefont {S.~F.~E.}\ \bibnamefont {Oliviero}},\ and\ \bibinfo {author} {\bibfnamefont {A.}~\bibnamefont {Hamma}},\ }\href {https://doi.org/10.1103/PhysRevLett.128.050402} {\bibfield  {journal} {\bibinfo  {journal} {Phys. Rev. Lett.}\ }\textbf {\bibinfo {volume} {128}},\ \bibinfo {pages} {050402} (\bibinfo {year} {2022})}\BibitemShut {NoStop}%
\bibitem [{\citenamefont {Leone}\ \emph {et~al.}(2023{\natexlab{a}})\citenamefont {Leone}, \citenamefont {Oliviero}, \citenamefont {Esposito},\ and\ \citenamefont {Hamma}}]{leone2023phase}%
  \BibitemOpen
  \bibfield  {author} {\bibinfo {author} {\bibfnamefont {L.}~\bibnamefont {Leone}}, \bibinfo {author} {\bibfnamefont {S.~F.~E.}\ \bibnamefont {Oliviero}}, \bibinfo {author} {\bibfnamefont {G.}~\bibnamefont {Esposito}},\ and\ \bibinfo {author} {\bibfnamefont {A.}~\bibnamefont {Hamma}},\ }\href@noop {} {} (\bibinfo {year} {2023}{\natexlab{a}}),\ \Eprint {https://arxiv.org/abs/2302.07895} {arXiv:2302.07895} \BibitemShut {NoStop}%
\bibitem [{\citenamefont {Oliviero}\ \emph {et~al.}(2021)\citenamefont {Oliviero}, \citenamefont {Leone},\ and\ \citenamefont {Hamma}}]{oliviero2021transitionsinentanglementcomplexity}%
  \BibitemOpen
  \bibfield  {author} {\bibinfo {author} {\bibfnamefont {S.~F.}\ \bibnamefont {Oliviero}}, \bibinfo {author} {\bibfnamefont {L.}~\bibnamefont {Leone}},\ and\ \bibinfo {author} {\bibfnamefont {A.}~\bibnamefont {Hamma}},\ }\href {https://doi.org/10.1016/j.physleta.2021.127721} {\bibfield  {journal} {\bibinfo  {journal} {Physics Letters A}\ }\textbf {\bibinfo {volume} {418}},\ \bibinfo {pages} {127721} (\bibinfo {year} {2021})}\BibitemShut {NoStop}%
\bibitem [{\citenamefont {Oliviero}\ \emph {et~al.}(2022{\natexlab{a}})\citenamefont {Oliviero}, \citenamefont {Leone},\ and\ \citenamefont {Hamma}}]{oliviero2022magicstateresourcetheory}%
  \BibitemOpen
  \bibfield  {author} {\bibinfo {author} {\bibfnamefont {S.~F.~E.}\ \bibnamefont {Oliviero}}, \bibinfo {author} {\bibfnamefont {L.}~\bibnamefont {Leone}},\ and\ \bibinfo {author} {\bibfnamefont {A.}~\bibnamefont {Hamma}},\ }\href {https://doi.org/10.1103/PhysRevA.106.042426} {\bibfield  {journal} {\bibinfo  {journal} {Phys. Rev. A}\ }\textbf {\bibinfo {volume} {106}},\ \bibinfo {pages} {042426} (\bibinfo {year} {2022}{\natexlab{a}})}\BibitemShut {NoStop}%
\bibitem [{\citenamefont {Tirrito}\ \emph {et~al.}()\citenamefont {Tirrito}, \citenamefont {Tarabunga}, \citenamefont {Lami}, \citenamefont {Chanda}, \citenamefont {Leone}, \citenamefont {Oliviero}, \citenamefont {Dalmonte}, \citenamefont {Collura},\ and\ \citenamefont {Hamma}}]{tirrito2023quantifying}%
  \BibitemOpen
  \bibfield  {author} {\bibinfo {author} {\bibfnamefont {E.}~\bibnamefont {Tirrito}}, \bibinfo {author} {\bibfnamefont {P.~S.}\ \bibnamefont {Tarabunga}}, \bibinfo {author} {\bibfnamefont {G.}~\bibnamefont {Lami}}, \bibinfo {author} {\bibfnamefont {T.}~\bibnamefont {Chanda}}, \bibinfo {author} {\bibfnamefont {L.}~\bibnamefont {Leone}}, \bibinfo {author} {\bibfnamefont {S.~F.~E.}\ \bibnamefont {Oliviero}}, \bibinfo {author} {\bibfnamefont {M.}~\bibnamefont {Dalmonte}}, \bibinfo {author} {\bibfnamefont {M.}~\bibnamefont {Collura}},\ and\ \bibinfo {author} {\bibfnamefont {A.}~\bibnamefont {Hamma}},\ }\href@noop {} {}\Eprint {https://arxiv.org/abs/2304.01175} {arXiv:2304.01175} \BibitemShut {NoStop}%
\bibitem [{\citenamefont {Oliviero}\ \emph {et~al.}(2022{\natexlab{b}})\citenamefont {Oliviero}, \citenamefont {Leone}, \citenamefont {Hamma},\ and\ \citenamefont {Lloyd}}]{oliviero2022measuringmagicon}%
  \BibitemOpen
  \bibfield  {author} {\bibinfo {author} {\bibfnamefont {S.~F.~E.}\ \bibnamefont {Oliviero}}, \bibinfo {author} {\bibfnamefont {L.}~\bibnamefont {Leone}}, \bibinfo {author} {\bibfnamefont {A.}~\bibnamefont {Hamma}},\ and\ \bibinfo {author} {\bibfnamefont {S.}~\bibnamefont {Lloyd}},\ }\href {https://doi.org/10.1038/s41534-022-00666-5} {\bibfield  {journal} {\bibinfo  {journal} {npj Quantum Inf.}\ }\textbf {\bibinfo {volume} {8}} (\bibinfo {year} {2022}{\natexlab{b}})}\BibitemShut {NoStop}%
\bibitem [{\citenamefont {Haug}\ and\ \citenamefont {Kim}(2023)}]{haug2023scalabremeasuresof}%
  \BibitemOpen
  \bibfield  {author} {\bibinfo {author} {\bibfnamefont {T.}~\bibnamefont {Haug}}\ and\ \bibinfo {author} {\bibfnamefont {M.}~\bibnamefont {Kim}},\ }\href {https://doi.org/10.1103/PRXQuantum.4.010301} {\bibfield  {journal} {\bibinfo  {journal} {PRX Quantum}\ }\textbf {\bibinfo {volume} {4}},\ \bibinfo {pages} {010301} (\bibinfo {year} {2023})}\BibitemShut {NoStop}%
\bibitem [{\citenamefont {Haug}\ and\ \citenamefont {Piroli}(2023)}]{haug2023quantifyingnonstabilizernessof}%
  \BibitemOpen
  \bibfield  {author} {\bibinfo {author} {\bibfnamefont {T.}~\bibnamefont {Haug}}\ and\ \bibinfo {author} {\bibfnamefont {L.}~\bibnamefont {Piroli}},\ }\href {https://doi.org/10.1103/PhysRevB.107.035148} {\bibfield  {journal} {\bibinfo  {journal} {Phys. Rev. B}\ }\textbf {\bibinfo {volume} {107}},\ \bibinfo {pages} {035148} (\bibinfo {year} {2023})}\BibitemShut {NoStop}%
\bibitem [{\citenamefont {Lami}\ and\ \citenamefont {Collura}()}]{lami2023quantum}%
  \BibitemOpen
  \bibfield  {author} {\bibinfo {author} {\bibfnamefont {G.}~\bibnamefont {Lami}}\ and\ \bibinfo {author} {\bibfnamefont {M.}~\bibnamefont {Collura}},\ }\href@noop {} {}\Eprint {https://arxiv.org/abs/2303.05536} {arXiv:2303.05536} \BibitemShut {NoStop}%
\bibitem [{\citenamefont {Castellani}\ and\ \citenamefont {Peliti}(1986)}]{castellani1986multifractal}%
  \BibitemOpen
  \bibfield  {author} {\bibinfo {author} {\bibfnamefont {C.}~\bibnamefont {Castellani}}\ and\ \bibinfo {author} {\bibfnamefont {L.}~\bibnamefont {Peliti}},\ }\href {https://doi.org/10.1088/0305-4470/19/8/004} {\bibfield  {journal} {\bibinfo  {journal} {J. Phys. A: Math. Theor.}\ }\textbf {\bibinfo {volume} {19}},\ \bibinfo {pages} {L429} (\bibinfo {year} {1986})}\BibitemShut {NoStop}%
\bibitem [{\citenamefont {Evers}\ and\ \citenamefont {Mirlin}(2000)}]{evers2000fluctuationsofthe}%
  \BibitemOpen
  \bibfield  {author} {\bibinfo {author} {\bibfnamefont {F.}~\bibnamefont {Evers}}\ and\ \bibinfo {author} {\bibfnamefont {A.~D.}\ \bibnamefont {Mirlin}},\ }\href {https://doi.org/10.1103/PhysRevLett.84.3690} {\bibfield  {journal} {\bibinfo  {journal} {Phys. Rev. Lett.}\ }\textbf {\bibinfo {volume} {84}},\ \bibinfo {pages} {3690} (\bibinfo {year} {2000})}\BibitemShut {NoStop}%
\bibitem [{\citenamefont {Evers}\ and\ \citenamefont {Mirlin}(2008)}]{evers2008andersontransitions}%
  \BibitemOpen
  \bibfield  {author} {\bibinfo {author} {\bibfnamefont {F.}~\bibnamefont {Evers}}\ and\ \bibinfo {author} {\bibfnamefont {A.~D.}\ \bibnamefont {Mirlin}},\ }\href {https://doi.org/10.1103/RevModPhys.80.1355} {\bibfield  {journal} {\bibinfo  {journal} {Rev. Mod. Phys.}\ }\textbf {\bibinfo {volume} {80}},\ \bibinfo {pages} {1355} (\bibinfo {year} {2008})}\BibitemShut {NoStop}%
\bibitem [{\citenamefont {Rodriguez}\ \emph {et~al.}(2009)\citenamefont {Rodriguez}, \citenamefont {Vasquez},\ and\ \citenamefont {R\"omer}}]{rodriguez2009}%
  \BibitemOpen
  \bibfield  {author} {\bibinfo {author} {\bibfnamefont {A.}~\bibnamefont {Rodriguez}}, \bibinfo {author} {\bibfnamefont {L.~J.}\ \bibnamefont {Vasquez}},\ and\ \bibinfo {author} {\bibfnamefont {R.~A.}\ \bibnamefont {R\"omer}},\ }\href {https://doi.org/10.1103/PhysRevLett.102.106406} {\bibfield  {journal} {\bibinfo  {journal} {Phys. Rev. Lett.}\ }\textbf {\bibinfo {volume} {102}},\ \bibinfo {pages} {106406} (\bibinfo {year} {2009})}\BibitemShut {NoStop}%
\bibitem [{\citenamefont {Rodriguez}\ \emph {et~al.}(2010)\citenamefont {Rodriguez}, \citenamefont {Vasquez}, \citenamefont {Slevin},\ and\ \citenamefont {R\"omer}}]{rodriguez2010critical}%
  \BibitemOpen
  \bibfield  {author} {\bibinfo {author} {\bibfnamefont {A.}~\bibnamefont {Rodriguez}}, \bibinfo {author} {\bibfnamefont {L.~J.}\ \bibnamefont {Vasquez}}, \bibinfo {author} {\bibfnamefont {K.}~\bibnamefont {Slevin}},\ and\ \bibinfo {author} {\bibfnamefont {R.~A.}\ \bibnamefont {R\"omer}},\ }\href {https://doi.org/10.1103/PhysRevLett.105.046403} {\bibfield  {journal} {\bibinfo  {journal} {Phys. Rev. Lett.}\ }\textbf {\bibinfo {volume} {105}},\ \bibinfo {pages} {046403} (\bibinfo {year} {2010})}\BibitemShut {NoStop}%
\bibitem [{\citenamefont {{De Luca}}\ and\ \citenamefont {Scardicchio}(2013)}]{DeLuca2013}%
  \BibitemOpen
  \bibfield  {author} {\bibinfo {author} {\bibfnamefont {A.}~\bibnamefont {{De Luca}}}\ and\ \bibinfo {author} {\bibfnamefont {A.}~\bibnamefont {Scardicchio}},\ }\href {https://doi.org/10.1209/0295-5075/101/37003} {\bibfield  {journal} {\bibinfo  {journal} {Europhys. Lett.}\ }\textbf {\bibinfo {volume} {101}},\ \bibinfo {pages} {37003} (\bibinfo {year} {2013})}\BibitemShut {NoStop}%
\bibitem [{\citenamefont {Sierant}\ \emph {et~al.}()\citenamefont {Sierant}, \citenamefont {Lewenstein},\ and\ \citenamefont {Scardicchio}}]{sierant2023universality}%
  \BibitemOpen
  \bibfield  {author} {\bibinfo {author} {\bibfnamefont {P.}~\bibnamefont {Sierant}}, \bibinfo {author} {\bibfnamefont {M.}~\bibnamefont {Lewenstein}},\ and\ \bibinfo {author} {\bibfnamefont {A.}~\bibnamefont {Scardicchio}},\ }\href@noop {} {}\Eprint {https://arxiv.org/abs/2205.14614} {arXiv:2205.14614} \BibitemShut {NoStop}%
\bibitem [{\citenamefont {St\'ephan}\ \emph {et~al.}(2009)\citenamefont {St\'ephan}, \citenamefont {Furukawa}, \citenamefont {Misguich},\ and\ \citenamefont {Pasquier}}]{stephan2009}%
  \BibitemOpen
  \bibfield  {author} {\bibinfo {author} {\bibfnamefont {J.-M.}\ \bibnamefont {St\'ephan}}, \bibinfo {author} {\bibfnamefont {S.}~\bibnamefont {Furukawa}}, \bibinfo {author} {\bibfnamefont {G.}~\bibnamefont {Misguich}},\ and\ \bibinfo {author} {\bibfnamefont {V.}~\bibnamefont {Pasquier}},\ }\href {https://doi.org/10.1103/PhysRevB.80.184421} {\bibfield  {journal} {\bibinfo  {journal} {Phys. Rev. B}\ }\textbf {\bibinfo {volume} {80}},\ \bibinfo {pages} {184421} (\bibinfo {year} {2009})}\BibitemShut {NoStop}%
\bibitem [{\citenamefont {St\'ephan}\ \emph {et~al.}(2010)\citenamefont {St\'ephan}, \citenamefont {Misguich},\ and\ \citenamefont {Pasquier}}]{stephan2010}%
  \BibitemOpen
  \bibfield  {author} {\bibinfo {author} {\bibfnamefont {J.-M.}\ \bibnamefont {St\'ephan}}, \bibinfo {author} {\bibfnamefont {G.}~\bibnamefont {Misguich}},\ and\ \bibinfo {author} {\bibfnamefont {V.}~\bibnamefont {Pasquier}},\ }\href {https://doi.org/10.1103/PhysRevB.82.125455} {\bibfield  {journal} {\bibinfo  {journal} {Phys. Rev. B}\ }\textbf {\bibinfo {volume} {82}},\ \bibinfo {pages} {125455} (\bibinfo {year} {2010})}\BibitemShut {NoStop}%
\bibitem [{\citenamefont {St\'ephan}(2014)}]{stephan2014shannon}%
  \BibitemOpen
  \bibfield  {author} {\bibinfo {author} {\bibfnamefont {J.-M.}\ \bibnamefont {St\'ephan}},\ }\href {https://doi.org/10.1103/PhysRevB.90.045424} {\bibfield  {journal} {\bibinfo  {journal} {Phys. Rev. B}\ }\textbf {\bibinfo {volume} {90}},\ \bibinfo {pages} {045424} (\bibinfo {year} {2014})}\BibitemShut {NoStop}%
\bibitem [{\citenamefont {Luitz}\ \emph {et~al.}(2014{\natexlab{a}})\citenamefont {Luitz}, \citenamefont {Alet},\ and\ \citenamefont {Laflorencie}}]{luitz2014universalbehaviorbeyond}%
  \BibitemOpen
  \bibfield  {author} {\bibinfo {author} {\bibfnamefont {D.~J.}\ \bibnamefont {Luitz}}, \bibinfo {author} {\bibfnamefont {F.}~\bibnamefont {Alet}},\ and\ \bibinfo {author} {\bibfnamefont {N.}~\bibnamefont {Laflorencie}},\ }\href {https://doi.org/10.1103/PhysRevLett.112.057203} {\bibfield  {journal} {\bibinfo  {journal} {Phys. Rev. Lett.}\ }\textbf {\bibinfo {volume} {112}},\ \bibinfo {pages} {057203} (\bibinfo {year} {2014}{\natexlab{a}})}\BibitemShut {NoStop}%
\bibitem [{\citenamefont {Fradkin}\ and\ \citenamefont {Moore}(2006)}]{fradkin2006entanglemententropyof}%
  \BibitemOpen
  \bibfield  {author} {\bibinfo {author} {\bibfnamefont {E.}~\bibnamefont {Fradkin}}\ and\ \bibinfo {author} {\bibfnamefont {J.~E.}\ \bibnamefont {Moore}},\ }\href {https://doi.org/10.1103/PhysRevLett.97.050404} {\bibfield  {journal} {\bibinfo  {journal} {Phys. Rev. Lett.}\ }\textbf {\bibinfo {volume} {97}},\ \bibinfo {pages} {050404} (\bibinfo {year} {2006})}\BibitemShut {NoStop}%
\bibitem [{\citenamefont {Zaletel}\ \emph {et~al.}(2011)\citenamefont {Zaletel}, \citenamefont {Bardarson},\ and\ \citenamefont {Moore}}]{zaletel2011logarithmicterms}%
  \BibitemOpen
  \bibfield  {author} {\bibinfo {author} {\bibfnamefont {M.~P.}\ \bibnamefont {Zaletel}}, \bibinfo {author} {\bibfnamefont {J.~H.}\ \bibnamefont {Bardarson}},\ and\ \bibinfo {author} {\bibfnamefont {J.~E.}\ \bibnamefont {Moore}},\ }\href {https://doi.org/10.1103/PhysRevLett.107.020402} {\bibfield  {journal} {\bibinfo  {journal} {Phys. Rev. Lett.}\ }\textbf {\bibinfo {volume} {107}},\ \bibinfo {pages} {020402} (\bibinfo {year} {2011})}\BibitemShut {NoStop}%
\bibitem [{\citenamefont {Alcaraz}\ and\ \citenamefont {Rajabpour}(2013)}]{alcaraz2013universalbehaviorof}%
  \BibitemOpen
  \bibfield  {author} {\bibinfo {author} {\bibfnamefont {F.~C.}\ \bibnamefont {Alcaraz}}\ and\ \bibinfo {author} {\bibfnamefont {M.~A.}\ \bibnamefont {Rajabpour}},\ }\href {https://doi.org/10.1103/PhysRevLett.111.017201} {\bibfield  {journal} {\bibinfo  {journal} {Phys. Rev. Lett.}\ }\textbf {\bibinfo {volume} {111}},\ \bibinfo {pages} {017201} (\bibinfo {year} {2013})}\BibitemShut {NoStop}%
\bibitem [{\citenamefont {Lindinger}\ \emph {et~al.}(2019)\citenamefont {Lindinger}, \citenamefont {Buchleitner},\ and\ \citenamefont {Rodr\'{\i}guez}}]{lindinger2019manybodymultifractality}%
  \BibitemOpen
  \bibfield  {author} {\bibinfo {author} {\bibfnamefont {J.}~\bibnamefont {Lindinger}}, \bibinfo {author} {\bibfnamefont {A.}~\bibnamefont {Buchleitner}},\ and\ \bibinfo {author} {\bibfnamefont {A.}~\bibnamefont {Rodr\'{\i}guez}},\ }\href {https://doi.org/10.1103/PhysRevLett.122.106603} {\bibfield  {journal} {\bibinfo  {journal} {Phys. Rev. Lett.}\ }\textbf {\bibinfo {volume} {122}},\ \bibinfo {pages} {106603} (\bibinfo {year} {2019})}\BibitemShut {NoStop}%
\bibitem [{\citenamefont {Luitz}\ \emph {et~al.}(2014{\natexlab{b}})\citenamefont {Luitz}, \citenamefont {Alet},\ and\ \citenamefont {Laflorencie}}]{luitz2014shannonrenyientropy}%
  \BibitemOpen
  \bibfield  {author} {\bibinfo {author} {\bibfnamefont {D.~J.}\ \bibnamefont {Luitz}}, \bibinfo {author} {\bibfnamefont {F.}~\bibnamefont {Alet}},\ and\ \bibinfo {author} {\bibfnamefont {N.}~\bibnamefont {Laflorencie}},\ }\href {https://doi.org/10.1103/PhysRevB.89.165106} {\bibfield  {journal} {\bibinfo  {journal} {Phys. Rev. B}\ }\textbf {\bibinfo {volume} {89}},\ \bibinfo {pages} {165106} (\bibinfo {year} {2014}{\natexlab{b}})}\BibitemShut {NoStop}%
\bibitem [{\citenamefont {Atas}\ and\ \citenamefont {Bogomolny}(2012)}]{atas2012multifractality}%
  \BibitemOpen
  \bibfield  {author} {\bibinfo {author} {\bibfnamefont {Y.~Y.}\ \bibnamefont {Atas}}\ and\ \bibinfo {author} {\bibfnamefont {E.}~\bibnamefont {Bogomolny}},\ }\href {https://doi.org/10.1103/PhysRevE.86.021104} {\bibfield  {journal} {\bibinfo  {journal} {Phys. Rev. E}\ }\textbf {\bibinfo {volume} {86}},\ \bibinfo {pages} {021104} (\bibinfo {year} {2012})}\BibitemShut {NoStop}%
\bibitem [{\citenamefont {Luitz}\ \emph {et~al.}(2014{\natexlab{c}})\citenamefont {Luitz}, \citenamefont {Laflorencie},\ and\ \citenamefont {Alet}}]{Luitz2014}%
  \BibitemOpen
  \bibfield  {author} {\bibinfo {author} {\bibfnamefont {D.~J.}\ \bibnamefont {Luitz}}, \bibinfo {author} {\bibfnamefont {N.}~\bibnamefont {Laflorencie}},\ and\ \bibinfo {author} {\bibfnamefont {F.}~\bibnamefont {Alet}},\ }\href {https://doi.org/10.1088/1742-5468/2014/08/p08007} {\bibfield  {journal} {\bibinfo  {journal} {J. Stat. Mech. Theory Exp.}\ }\textbf {\bibinfo {volume} {2014}},\ \bibinfo {pages} {P08007} (\bibinfo {year} {2014}{\natexlab{c}})}\BibitemShut {NoStop}%
\bibitem [{\citenamefont {Luitz}\ \emph {et~al.}(2020)\citenamefont {Luitz}, \citenamefont {Khaymovich},\ and\ \citenamefont {Lev}}]{luitz2020multifractality}%
  \BibitemOpen
  \bibfield  {author} {\bibinfo {author} {\bibfnamefont {D.~J.}\ \bibnamefont {Luitz}}, \bibinfo {author} {\bibfnamefont {I.~M.}\ \bibnamefont {Khaymovich}},\ and\ \bibinfo {author} {\bibfnamefont {Y.~B.}\ \bibnamefont {Lev}},\ }\href {https://doi.org/10.21468/SciPostPhysCore.2.2.006} {\bibfield  {journal} {\bibinfo  {journal} {SciPost Phys. Core}\ }\textbf {\bibinfo {volume} {2}},\ \bibinfo {pages} {006} (\bibinfo {year} {2020})}\BibitemShut {NoStop}%
\bibitem [{\citenamefont {Mac\'e}\ \emph {et~al.}(2019)\citenamefont {Mac\'e}, \citenamefont {Alet},\ and\ \citenamefont {Laflorencie}}]{mace2019multifractal}%
  \BibitemOpen
  \bibfield  {author} {\bibinfo {author} {\bibfnamefont {N.}~\bibnamefont {Mac\'e}}, \bibinfo {author} {\bibfnamefont {F.}~\bibnamefont {Alet}},\ and\ \bibinfo {author} {\bibfnamefont {N.}~\bibnamefont {Laflorencie}},\ }\href {https://doi.org/10.1103/PhysRevLett.123.180601} {\bibfield  {journal} {\bibinfo  {journal} {Phys. Rev. Lett.}\ }\textbf {\bibinfo {volume} {123}},\ \bibinfo {pages} {180601} (\bibinfo {year} {2019})}\BibitemShut {NoStop}%
\bibitem [{\citenamefont {Tarzia}(2020)}]{tarzia2020manybody}%
  \BibitemOpen
  \bibfield  {author} {\bibinfo {author} {\bibfnamefont {M.}~\bibnamefont {Tarzia}},\ }\href {https://doi.org/10.1103/PhysRevB.102.014208} {\bibfield  {journal} {\bibinfo  {journal} {Phys. Rev. B}\ }\textbf {\bibinfo {volume} {102}},\ \bibinfo {pages} {014208} (\bibinfo {year} {2020})}\BibitemShut {NoStop}%
\bibitem [{\citenamefont {Sol\'orzano}\ \emph {et~al.}(2021)\citenamefont {Sol\'orzano}, \citenamefont {Santos},\ and\ \citenamefont {Torres-Herrera}}]{solorzano2021}%
  \BibitemOpen
  \bibfield  {author} {\bibinfo {author} {\bibfnamefont {A.}~\bibnamefont {Sol\'orzano}}, \bibinfo {author} {\bibfnamefont {L.~F.}\ \bibnamefont {Santos}},\ and\ \bibinfo {author} {\bibfnamefont {E.~J.}\ \bibnamefont {Torres-Herrera}},\ }\href {https://doi.org/10.1103/PhysRevResearch.3.L032030} {\bibfield  {journal} {\bibinfo  {journal} {Phys. Rev. Res.}\ }\textbf {\bibinfo {volume} {3}},\ \bibinfo {pages} {L032030} (\bibinfo {year} {2021})}\BibitemShut {NoStop}%
\bibitem [{\citenamefont {Pietracaprina}\ and\ \citenamefont {Laflorencie}(2021)}]{Pietracaprina2021}%
  \BibitemOpen
  \bibfield  {author} {\bibinfo {author} {\bibfnamefont {F.}~\bibnamefont {Pietracaprina}}\ and\ \bibinfo {author} {\bibfnamefont {N.}~\bibnamefont {Laflorencie}},\ }\href {https://doi.org/10.1016/j.aop.2021.168502} {\bibfield  {journal} {\bibinfo  {journal} {Ann. Phys.}\ }\textbf {\bibinfo {volume} {435}},\ \bibinfo {pages} {168502} (\bibinfo {year} {2021})}\BibitemShut {NoStop}%
\bibitem [{\citenamefont {Monthus}(2016)}]{Monthus2016}%
  \BibitemOpen
  \bibfield  {author} {\bibinfo {author} {\bibfnamefont {C.}~\bibnamefont {Monthus}},\ }\href {https://doi.org/10.1088/1742-5468/2016/07/073301} {\bibfield  {journal} {\bibinfo  {journal} {J. Stat. Mech. Theory Exp.}\ }\textbf {\bibinfo {volume} {2016}},\ \bibinfo {pages} {073301} (\bibinfo {year} {2016})}\BibitemShut {NoStop}%
\bibitem [{\citenamefont {Pausch}\ \emph {et~al.}(2021)\citenamefont {Pausch}, \citenamefont {Carnio}, \citenamefont {Rodr\'{\i}guez},\ and\ \citenamefont {Buchleitner}}]{pausch2021chaosanderdoficity}%
  \BibitemOpen
  \bibfield  {author} {\bibinfo {author} {\bibfnamefont {L.}~\bibnamefont {Pausch}}, \bibinfo {author} {\bibfnamefont {E.~G.}\ \bibnamefont {Carnio}}, \bibinfo {author} {\bibfnamefont {A.}~\bibnamefont {Rodr\'{\i}guez}},\ and\ \bibinfo {author} {\bibfnamefont {A.}~\bibnamefont {Buchleitner}},\ }\href {https://doi.org/10.1103/PhysRevLett.126.150601} {\bibfield  {journal} {\bibinfo  {journal} {Phys. Rev. Lett.}\ }\textbf {\bibinfo {volume} {126}},\ \bibinfo {pages} {150601} (\bibinfo {year} {2021})}\BibitemShut {NoStop}%
\bibitem [{\citenamefont {Sierant}\ \emph {et~al.}(2023)\citenamefont {Sierant}, \citenamefont {Lewenstein}, \citenamefont {Scardicchio},\ and\ \citenamefont {Zakrzewski}}]{sierant2023stability}%
  \BibitemOpen
  \bibfield  {author} {\bibinfo {author} {\bibfnamefont {P.}~\bibnamefont {Sierant}}, \bibinfo {author} {\bibfnamefont {M.}~\bibnamefont {Lewenstein}}, \bibinfo {author} {\bibfnamefont {A.}~\bibnamefont {Scardicchio}},\ and\ \bibinfo {author} {\bibfnamefont {J.}~\bibnamefont {Zakrzewski}},\ }\href {https://doi.org/10.1103/PhysRevB.107.115132} {\bibfield  {journal} {\bibinfo  {journal} {Phys. Rev. B}\ }\textbf {\bibinfo {volume} {107}},\ \bibinfo {pages} {115132} (\bibinfo {year} {2023})}\BibitemShut {NoStop}%
\bibitem [{\citenamefont {De~Tomasi}\ and\ \citenamefont {Khaymovich}(2020)}]{detomasi2020multifractality}%
  \BibitemOpen
  \bibfield  {author} {\bibinfo {author} {\bibfnamefont {G.}~\bibnamefont {De~Tomasi}}\ and\ \bibinfo {author} {\bibfnamefont {I.~M.}\ \bibnamefont {Khaymovich}},\ }\href {https://doi.org/10.1103/PhysRevLett.124.200602} {\bibfield  {journal} {\bibinfo  {journal} {Phys. Rev. Lett.}\ }\textbf {\bibinfo {volume} {124}},\ \bibinfo {pages} {200602} (\bibinfo {year} {2020})}\BibitemShut {NoStop}%
\bibitem [{\citenamefont {Roy}\ and\ \citenamefont {Logan}(2021)}]{roy2021fockspace}%
  \BibitemOpen
  \bibfield  {author} {\bibinfo {author} {\bibfnamefont {S.}~\bibnamefont {Roy}}\ and\ \bibinfo {author} {\bibfnamefont {D.~E.}\ \bibnamefont {Logan}},\ }\href {https://doi.org/10.1103/PhysRevB.104.174201} {\bibfield  {journal} {\bibinfo  {journal} {Phys. Rev. B}\ }\textbf {\bibinfo {volume} {104}},\ \bibinfo {pages} {174201} (\bibinfo {year} {2021})}\BibitemShut {NoStop}%
\bibitem [{\citenamefont {De~Tomasi}\ \emph {et~al.}(2021)\citenamefont {De~Tomasi}, \citenamefont {Khaymovich}, \citenamefont {Pollmann},\ and\ \citenamefont {Warzel}}]{detomasi2021rare}%
  \BibitemOpen
  \bibfield  {author} {\bibinfo {author} {\bibfnamefont {G.}~\bibnamefont {De~Tomasi}}, \bibinfo {author} {\bibfnamefont {I.~M.}\ \bibnamefont {Khaymovich}}, \bibinfo {author} {\bibfnamefont {F.}~\bibnamefont {Pollmann}},\ and\ \bibinfo {author} {\bibfnamefont {S.}~\bibnamefont {Warzel}},\ }\href {https://doi.org/10.1103/PhysRevB.104.024202} {\bibfield  {journal} {\bibinfo  {journal} {Phys. Rev. B}\ }\textbf {\bibinfo {volume} {104}},\ \bibinfo {pages} {024202} (\bibinfo {year} {2021})}\BibitemShut {NoStop}%
\bibitem [{\citenamefont {Biroli}\ \emph {et~al.}(2021)\citenamefont {Biroli}, \citenamefont {Facoetti}, \citenamefont {Schir\'o}, \citenamefont {Tarzia},\ and\ \citenamefont {Vivo}}]{biroli2021outofequilibrium}%
  \BibitemOpen
  \bibfield  {author} {\bibinfo {author} {\bibfnamefont {G.}~\bibnamefont {Biroli}}, \bibinfo {author} {\bibfnamefont {D.}~\bibnamefont {Facoetti}}, \bibinfo {author} {\bibfnamefont {M.}~\bibnamefont {Schir\'o}}, \bibinfo {author} {\bibfnamefont {M.}~\bibnamefont {Tarzia}},\ and\ \bibinfo {author} {\bibfnamefont {P.}~\bibnamefont {Vivo}},\ }\href {https://doi.org/10.1103/PhysRevB.103.014204} {\bibfield  {journal} {\bibinfo  {journal} {Phys. Rev. B}\ }\textbf {\bibinfo {volume} {103}},\ \bibinfo {pages} {014204} (\bibinfo {year} {2021})}\BibitemShut {NoStop}%
\bibitem [{\citenamefont {Roy}(2022)}]{roy2022hilbertspace}%
  \BibitemOpen
  \bibfield  {author} {\bibinfo {author} {\bibfnamefont {S.}~\bibnamefont {Roy}},\ }\href {https://doi.org/10.1103/PhysRevB.106.L140204} {\bibfield  {journal} {\bibinfo  {journal} {Phys. Rev. B}\ }\textbf {\bibinfo {volume} {106}},\ \bibinfo {pages} {L140204} (\bibinfo {year} {2022})}\BibitemShut {NoStop}%
\bibitem [{\citenamefont {Sierant}\ and\ \citenamefont {Turkeshi}(2022)}]{sierant2022universalbehaviorbeyond}%
  \BibitemOpen
  \bibfield  {author} {\bibinfo {author} {\bibfnamefont {P.}~\bibnamefont {Sierant}}\ and\ \bibinfo {author} {\bibfnamefont {X.}~\bibnamefont {Turkeshi}},\ }\href {https://doi.org/10.1103/PhysRevLett.128.130605} {\bibfield  {journal} {\bibinfo  {journal} {Phys. Rev. Lett.}\ }\textbf {\bibinfo {volume} {128}},\ \bibinfo {pages} {130605} (\bibinfo {year} {2022})}\BibitemShut {NoStop}%
\bibitem [{\citenamefont {Sierant}\ \emph {et~al.}(2022)\citenamefont {Sierant}, \citenamefont {Schir\`o}, \citenamefont {Lewenstein},\ and\ \citenamefont {Turkeshi}}]{Sierant22}%
  \BibitemOpen
  \bibfield  {author} {\bibinfo {author} {\bibfnamefont {P.}~\bibnamefont {Sierant}}, \bibinfo {author} {\bibfnamefont {M.}~\bibnamefont {Schir\`o}}, \bibinfo {author} {\bibfnamefont {M.}~\bibnamefont {Lewenstein}},\ and\ \bibinfo {author} {\bibfnamefont {X.}~\bibnamefont {Turkeshi}},\ }\href {https://doi.org/10.1103/PhysRevB.106.214316} {\bibfield  {journal} {\bibinfo  {journal} {Phys. Rev. B}\ }\textbf {\bibinfo {volume} {106}},\ \bibinfo {pages} {214316} (\bibinfo {year} {2022})}\BibitemShut {NoStop}%
\bibitem [{\citenamefont {Stanley}\ and\ \citenamefont {Meakin}(1988)}]{stanley1988multifractalphenomenain}%
  \BibitemOpen
  \bibfield  {author} {\bibinfo {author} {\bibfnamefont {H.~E.}\ \bibnamefont {Stanley}}\ and\ \bibinfo {author} {\bibfnamefont {P.}~\bibnamefont {Meakin}},\ }\href {https://doi.org/10.1038/335405a0} {\bibfield  {journal} {\bibinfo  {journal} {Nature}\ }\textbf {\bibinfo {volume} {335}},\ \bibinfo {pages} {405} (\bibinfo {year} {1988})}\BibitemShut {NoStop}%
\bibitem [{\citenamefont {Carrasco}\ \emph {et~al.}(2022)\citenamefont {Carrasco}, \citenamefont {Votto}, \citenamefont {Vitale}, \citenamefont {Kokail}, \citenamefont {Neven}, \citenamefont {Zoller}, \citenamefont {Vermersch},\ and\ \citenamefont {Kraus}}]{carrasco2022entanglement}%
  \BibitemOpen
  \bibfield  {author} {\bibinfo {author} {\bibfnamefont {J.}~\bibnamefont {Carrasco}}, \bibinfo {author} {\bibfnamefont {M.}~\bibnamefont {Votto}}, \bibinfo {author} {\bibfnamefont {V.}~\bibnamefont {Vitale}}, \bibinfo {author} {\bibfnamefont {C.}~\bibnamefont {Kokail}}, \bibinfo {author} {\bibfnamefont {A.}~\bibnamefont {Neven}}, \bibinfo {author} {\bibfnamefont {P.}~\bibnamefont {Zoller}}, \bibinfo {author} {\bibfnamefont {B.}~\bibnamefont {Vermersch}},\ and\ \bibinfo {author} {\bibfnamefont {B.}~\bibnamefont {Kraus}},\ }\href@noop {} {\bibinfo {title} {Entanglement phase diagrams from partial transpose moments}} (\bibinfo {year} {2022}),\ \Eprint {https://arxiv.org/abs/2212.10181} {arXiv:2212.10181 [quant-ph]} \BibitemShut {NoStop}%
\bibitem [{\citenamefont {Ferris}\ \emph {et~al.}(2022)\citenamefont {Ferris}, \citenamefont {Rasmusson}, \citenamefont {Bronn},\ and\ \citenamefont {Lanes}}]{ferris2022quantum}%
  \BibitemOpen
  \bibfield  {author} {\bibinfo {author} {\bibfnamefont {K.~J.}\ \bibnamefont {Ferris}}, \bibinfo {author} {\bibfnamefont {A.~J.}\ \bibnamefont {Rasmusson}}, \bibinfo {author} {\bibfnamefont {N.~T.}\ \bibnamefont {Bronn}},\ and\ \bibinfo {author} {\bibfnamefont {O.}~\bibnamefont {Lanes}},\ }\href@noop {} {} (\bibinfo {year} {2022}),\ \Eprint {https://arxiv.org/abs/2209.02795} {arXiv:2209.02795} \BibitemShut {NoStop}%
\bibitem [{\citenamefont {Preskill}(2018)}]{Preskill2018quantumcomputingin}%
  \BibitemOpen
  \bibfield  {author} {\bibinfo {author} {\bibfnamefont {J.}~\bibnamefont {Preskill}},\ }\href {https://doi.org/10.22331/q-2018-08-06-79} {\bibfield  {journal} {\bibinfo  {journal} {{Quantum}}\ }\textbf {\bibinfo {volume} {2}},\ \bibinfo {pages} {79} (\bibinfo {year} {2018})}\BibitemShut {NoStop}%
\bibitem [{\citenamefont {Fraxanet}\ \emph {et~al.}()\citenamefont {Fraxanet}, \citenamefont {Salamon},\ and\ \citenamefont {Lewenstein}}]{fraxanet2022coming}%
  \BibitemOpen
  \bibfield  {author} {\bibinfo {author} {\bibfnamefont {J.}~\bibnamefont {Fraxanet}}, \bibinfo {author} {\bibfnamefont {T.}~\bibnamefont {Salamon}},\ and\ \bibinfo {author} {\bibfnamefont {M.}~\bibnamefont {Lewenstein}},\ }\href@noop {} {}\Eprint {https://arxiv.org/abs/2204.08905} {arXiv:2204.08905} \BibitemShut {NoStop}%
\bibitem [{\citenamefont {B\"acker}\ \emph {et~al.}(2019)\citenamefont {B\"acker}, \citenamefont {Haque},\ and\ \citenamefont {Khaymovich}}]{backer2019multifractal}%
  \BibitemOpen
  \bibfield  {author} {\bibinfo {author} {\bibfnamefont {A.}~\bibnamefont {B\"acker}}, \bibinfo {author} {\bibfnamefont {M.}~\bibnamefont {Haque}},\ and\ \bibinfo {author} {\bibfnamefont {I.~M.}\ \bibnamefont {Khaymovich}},\ }\href {https://doi.org/10.1103/PhysRevE.100.032117} {\bibfield  {journal} {\bibinfo  {journal} {Phys. Rev. E}\ }\textbf {\bibinfo {volume} {100}},\ \bibinfo {pages} {032117} (\bibinfo {year} {2019})}\BibitemShut {NoStop}%
\bibitem [{Note1()}]{Note1}%
  \BibitemOpen
  \bibinfo {note} {It is also possible that $0<D_q<1$ and $D_{q_1}=D_{q_2}$ for all $q_1,q_2>0$. Such a state is called fractal.}\BibitemShut {Stop}%
\bibitem [{\citenamefont {Gidney}(2021)}]{gidney2021stimfaststabilizer}%
  \BibitemOpen
  \bibfield  {author} {\bibinfo {author} {\bibfnamefont {C.}~\bibnamefont {Gidney}},\ }\href {https://doi.org/10.22331/q-2021-07-06-497} {\bibfield  {journal} {\bibinfo  {journal} {{Quantum}}\ }\textbf {\bibinfo {volume} {5}},\ \bibinfo {pages} {497} (\bibinfo {year} {2021})}\BibitemShut {NoStop}%
\bibitem [{\citenamefont {Zhu}(2017)}]{zhu2017multiqubitcliffordgroups}%
  \BibitemOpen
  \bibfield  {author} {\bibinfo {author} {\bibfnamefont {H.}~\bibnamefont {Zhu}},\ }\href {https://doi.org/10.1103/PhysRevA.96.062336} {\bibfield  {journal} {\bibinfo  {journal} {Phys. Rev. A}\ }\textbf {\bibinfo {volume} {96}},\ \bibinfo {pages} {062336} (\bibinfo {year} {2017})}\BibitemShut {NoStop}%
\bibitem [{\citenamefont {Webb}(2016)}]{webb2016clifford}%
  \BibitemOpen
  \bibfield  {author} {\bibinfo {author} {\bibfnamefont {Z.}~\bibnamefont {Webb}},\ }\href@noop {} {\bibfield  {journal} {\bibinfo  {journal} {Quantum Inf Comput.}\ }\textbf {\bibinfo {volume} {16}},\ \bibinfo {pages} {1379} (\bibinfo {year} {2016})}\BibitemShut {NoStop}%
\bibitem [{\citenamefont {Zhu}\ \emph {et~al.}(2016)\citenamefont {Zhu}, \citenamefont {Kueng}, \citenamefont {Grassl},\ and\ \citenamefont {Gross}}]{zhu2016thecliffordgroup}%
  \BibitemOpen
  \bibfield  {author} {\bibinfo {author} {\bibfnamefont {H.}~\bibnamefont {Zhu}}, \bibinfo {author} {\bibfnamefont {R.}~\bibnamefont {Kueng}}, \bibinfo {author} {\bibfnamefont {M.}~\bibnamefont {Grassl}},\ and\ \bibinfo {author} {\bibfnamefont {D.}~\bibnamefont {Gross}},\ }\href@noop {} {} (\bibinfo {year} {2016}),\ \Eprint {https://arxiv.org/abs/1609.08172} {arXiv:1609.08172} \BibitemShut {NoStop}%
\bibitem [{\citenamefont {Nezami}\ and\ \citenamefont {Walter}(2020)}]{nezami2020multipartiteentanglementin}%
  \BibitemOpen
  \bibfield  {author} {\bibinfo {author} {\bibfnamefont {S.}~\bibnamefont {Nezami}}\ and\ \bibinfo {author} {\bibfnamefont {M.}~\bibnamefont {Walter}},\ }\href {https://doi.org/10.1103/PhysRevLett.125.241602} {\bibfield  {journal} {\bibinfo  {journal} {Phys. Rev. Lett.}\ }\textbf {\bibinfo {volume} {125}},\ \bibinfo {pages} {241602} (\bibinfo {year} {2020})}\BibitemShut {NoStop}%
\bibitem [{\citenamefont {Montealegre-Mora}\ and\ \citenamefont {Gross}()}]{montealegremora2022dualitytheoryfor}%
  \BibitemOpen
  \bibfield  {author} {\bibinfo {author} {\bibfnamefont {F.}~\bibnamefont {Montealegre-Mora}}\ and\ \bibinfo {author} {\bibfnamefont {D.}~\bibnamefont {Gross}},\ }\href@noop {} {}\Eprint {https://arxiv.org/abs/2208.01688} {arXiv:2208.01688} \BibitemShut {NoStop}%
\bibitem [{\citenamefont {Montealegre-Mora}\ and\ \citenamefont {Gross}(2021)}]{MontealegreMora2021}%
  \BibitemOpen
  \bibfield  {author} {\bibinfo {author} {\bibfnamefont {F.}~\bibnamefont {Montealegre-Mora}}\ and\ \bibinfo {author} {\bibfnamefont {D.}~\bibnamefont {Gross}},\ }\href {https://doi.org/10.1090/ert/563} {\bibfield  {journal} {\bibinfo  {journal} {Represent. Theory}\ }\textbf {\bibinfo {volume} {25}},\ \bibinfo {pages} {193} (\bibinfo {year} {2021})}\BibitemShut {NoStop}%
\bibitem [{\citenamefont {Bravyi}\ and\ \citenamefont {Maslov}(2021)}]{Bravyi_2021}%
  \BibitemOpen
  \bibfield  {author} {\bibinfo {author} {\bibfnamefont {S.}~\bibnamefont {Bravyi}}\ and\ \bibinfo {author} {\bibfnamefont {D.}~\bibnamefont {Maslov}},\ }\href {https://doi.org/10.1109%2Ftit.2021.3081415} {\bibfield  {journal} {\bibinfo  {journal} {{IEEE} Trans. Inf. Theory}\ }\textbf {\bibinfo {volume} {67}},\ \bibinfo {pages} {4546} (\bibinfo {year} {2021})}\BibitemShut {NoStop}%
\bibitem [{\citenamefont {Gross}\ \emph {et~al.}(2021)\citenamefont {Gross}, \citenamefont {Nezami},\ and\ \citenamefont {Walter}}]{Gross2021}%
  \BibitemOpen
  \bibfield  {author} {\bibinfo {author} {\bibfnamefont {D.}~\bibnamefont {Gross}}, \bibinfo {author} {\bibfnamefont {S.}~\bibnamefont {Nezami}},\ and\ \bibinfo {author} {\bibfnamefont {M.}~\bibnamefont {Walter}},\ }\href {https://doi.org/10.1007/s00220-021-04118-7} {\bibfield  {journal} {\bibinfo  {journal} {Commun. Math. Phys.}\ }\textbf {\bibinfo {volume} {385}},\ \bibinfo {pages} {1325} (\bibinfo {year} {2021})}\BibitemShut {NoStop}%
\bibitem [{\citenamefont {Goold}\ \emph {et~al.}(2016)\citenamefont {Goold}, \citenamefont {Huber}, \citenamefont {Riera}, \citenamefont {del Rio},\ and\ \citenamefont {Skrzypczyk}}]{Goold2016}%
  \BibitemOpen
  \bibfield  {author} {\bibinfo {author} {\bibfnamefont {J.}~\bibnamefont {Goold}}, \bibinfo {author} {\bibfnamefont {M.}~\bibnamefont {Huber}}, \bibinfo {author} {\bibfnamefont {A.}~\bibnamefont {Riera}}, \bibinfo {author} {\bibfnamefont {L.}~\bibnamefont {del Rio}},\ and\ \bibinfo {author} {\bibfnamefont {P.}~\bibnamefont {Skrzypczyk}},\ }\href {https://doi.org/10.1088/1751-8113/49/14/143001} {\bibfield  {journal} {\bibinfo  {journal} {J. Phys. A: Math. Theor.}\ }\textbf {\bibinfo {volume} {49}},\ \bibinfo {pages} {143001} (\bibinfo {year} {2016})}\BibitemShut {NoStop}%
\bibitem [{\citenamefont {Wilde}(2013)}]{Wilde2013}%
  \BibitemOpen
  \bibfield  {author} {\bibinfo {author} {\bibfnamefont {M.~M.}\ \bibnamefont {Wilde}},\ }\href {https://doi.org/10.1017/cbo9781139525343} {\emph {\bibinfo {title} {Quantum Information Theory}}}\ (\bibinfo  {publisher} {Cambridge University Press, Cambridge, England},\ \bibinfo {year} {2013})\BibitemShut {NoStop}%
\bibitem [{Note2()}]{Note2}%
  \BibitemOpen
  \bibinfo {note} {The exact expression is \begin {align*} & \protect \mathrm {std}_{U\in \protect \mathcal {U}(d)}(\protect \overline {\protect \mathcal {F}}[U|\Psi _0\rangle ])^2 = \\ &\protect \frac {8 \left (17 d^5+42 d^4-106 d^3-72 d^2+449 d-330\right )}{(d+1)^2 (d+2)^2 (d+3)^2 (d+4) (d+5) (d+6) (d+7)}. \end {align*} This formula requires evaluating generic correlators $\protect \mathbb {E}_{U\in \protect \mathcal {U}(d)}I_{q_1}^{r_1}[U|\Psi _0\rangle ]I_{q_2}^{r_2}[U|\Psi _0\rangle ]$ and result in expression similar to Eq.~\protect \eqref {eq:res} but more involved.}\BibitemShut {Stop}%
\bibitem [{Note3()}]{Note3}%
  \BibitemOpen
  \bibinfo {note} {We note that the resources for calculation of $M_2(\mathinner {|{\Psi }\rangle })$ directly according to the definition \protect \eqref {eq:se} scale as $O(2^{3N})$, while the cost of Monte Carlo sampling of the Clifford orbit $O(2^{2N})$ until a prescribed accuracy of $M_2(\mathinner {|{\Psi }\rangle })$ is achieved scales as $O(2^{2N})$. Nevertheless, the constant describing the scaling in the Monte Carlo sampling case is large, and direct numerical calculation of $M_2(\mathinner {|{\Psi }\rangle })$ according to \protect \eqref {eq:se} is more efficient at practically relevant system sizes $N\lesssim 14$.}\BibitemShut {Stop}%
\bibitem [{\citenamefont {{IBM Quantum. https://quantum-computing.ibm.com/, 2021}}()}]{ibmquantum}%
  \BibitemOpen
  \bibfield  {author} {\bibinfo {author} {\bibnamefont {{IBM Quantum. https://quantum-computing.ibm.com/, 2021}}},\ }\href@noop {} {}\BibitemShut {NoStop}%
\bibitem [{\citenamefont {Geller}(2020)}]{Geller20}%
  \BibitemOpen
  \bibfield  {author} {\bibinfo {author} {\bibfnamefont {M.~R.}\ \bibnamefont {Geller}},\ }\href {https://doi.org/10.1088/2058-9565/ab9591} {\bibfield  {journal} {\bibinfo  {journal} {Quantum Science and Technology}\ }\textbf {\bibinfo {volume} {5}},\ \bibinfo {pages} {03LT01} (\bibinfo {year} {2020})}\BibitemShut {NoStop}%
\bibitem [{\citenamefont {Giurgica-Tiron}\ \emph {et~al.}(2020)\citenamefont {Giurgica-Tiron}, \citenamefont {Hindy}, \citenamefont {LaRose}, \citenamefont {Mari},\ and\ \citenamefont {Zeng}}]{Giurgica20}%
  \BibitemOpen
  \bibfield  {author} {\bibinfo {author} {\bibfnamefont {T.}~\bibnamefont {Giurgica-Tiron}}, \bibinfo {author} {\bibfnamefont {Y.}~\bibnamefont {Hindy}}, \bibinfo {author} {\bibfnamefont {R.}~\bibnamefont {LaRose}}, \bibinfo {author} {\bibfnamefont {A.}~\bibnamefont {Mari}},\ and\ \bibinfo {author} {\bibfnamefont {W.~J.}\ \bibnamefont {Zeng}},\ }in\ \href {https://doi.org/10.1109/QCE49297.2020.00045} {\emph {\bibinfo {booktitle} {2020 IEEE International Conference on Quantum Computing and Engineering (QCE)}}}\ (\bibinfo {year} {2020})\ pp.\ \bibinfo {pages} {306--316}\BibitemShut {NoStop}%
\bibitem [{\citenamefont {Nation}\ \emph {et~al.}(2021)\citenamefont {Nation}, \citenamefont {Kang}, \citenamefont {Sundaresan},\ and\ \citenamefont {Gambetta}}]{Nation21}%
  \BibitemOpen
  \bibfield  {author} {\bibinfo {author} {\bibfnamefont {P.~D.}\ \bibnamefont {Nation}}, \bibinfo {author} {\bibfnamefont {H.}~\bibnamefont {Kang}}, \bibinfo {author} {\bibfnamefont {N.}~\bibnamefont {Sundaresan}},\ and\ \bibinfo {author} {\bibfnamefont {J.~M.}\ \bibnamefont {Gambetta}},\ }\href {https://doi.org/10.1103/PRXQuantum.2.040326} {\bibfield  {journal} {\bibinfo  {journal} {PRX Quantum}\ }\textbf {\bibinfo {volume} {2}},\ \bibinfo {pages} {040326} (\bibinfo {year} {2021})}\BibitemShut {NoStop}%
\bibitem [{\citenamefont {Smith}\ \emph {et~al.}(2021)\citenamefont {Smith}, \citenamefont {Khosla}, \citenamefont {Self},\ and\ \citenamefont {Kim}}]{Alistair21}%
  \BibitemOpen
  \bibfield  {author} {\bibinfo {author} {\bibfnamefont {A.~W.~R.}\ \bibnamefont {Smith}}, \bibinfo {author} {\bibfnamefont {K.~E.}\ \bibnamefont {Khosla}}, \bibinfo {author} {\bibfnamefont {C.~N.}\ \bibnamefont {Self}},\ and\ \bibinfo {author} {\bibfnamefont {M.~S.}\ \bibnamefont {Kim}},\ }\href {https://doi.org/10.1126/sciadv.abi8009} {\bibfield  {journal} {\bibinfo  {journal} {Science Advances}\ }\textbf {\bibinfo {volume} {7}},\ \bibinfo {pages} {eabi8009} (\bibinfo {year} {2021})},\ \Eprint {https://arxiv.org/abs/https://www.science.org/doi/pdf/10.1126/sciadv.abi8009} {https://www.science.org/doi/pdf/10.1126/sciadv.abi8009} \BibitemShut {NoStop}%
\bibitem [{\citenamefont {Hicks}\ \emph {et~al.}(2022)\citenamefont {Hicks}, \citenamefont {Kobrin}, \citenamefont {Bauer},\ and\ \citenamefont {Nachman}}]{Hicks22}%
  \BibitemOpen
  \bibfield  {author} {\bibinfo {author} {\bibfnamefont {R.}~\bibnamefont {Hicks}}, \bibinfo {author} {\bibfnamefont {B.}~\bibnamefont {Kobrin}}, \bibinfo {author} {\bibfnamefont {C.~W.}\ \bibnamefont {Bauer}},\ and\ \bibinfo {author} {\bibfnamefont {B.}~\bibnamefont {Nachman}},\ }\href {https://doi.org/10.1103/PhysRevA.105.012419} {\bibfield  {journal} {\bibinfo  {journal} {Phys. Rev. A}\ }\textbf {\bibinfo {volume} {105}},\ \bibinfo {pages} {012419} (\bibinfo {year} {2022})}\BibitemShut {NoStop}%
\bibitem [{\citenamefont {Leone}\ \emph {et~al.}()\citenamefont {Leone}, \citenamefont {Oliviero}, \citenamefont {Lloyd},\ and\ \citenamefont {Hamma}}]{leone2022learning}%
  \BibitemOpen
  \bibfield  {author} {\bibinfo {author} {\bibfnamefont {L.}~\bibnamefont {Leone}}, \bibinfo {author} {\bibfnamefont {S.~F.~E.}\ \bibnamefont {Oliviero}}, \bibinfo {author} {\bibfnamefont {S.}~\bibnamefont {Lloyd}},\ and\ \bibinfo {author} {\bibfnamefont {A.}~\bibnamefont {Hamma}},\ }\href@noop {} {}\Eprint {https://arxiv.org/abs/2212.11338} {arXiv:2212.11338} \BibitemShut {NoStop}%
\bibitem [{\citenamefont {Pappalardi}\ and\ \citenamefont {Kurchan}(2023)}]{Pappalardi2023}%
  \BibitemOpen
  \bibfield  {author} {\bibinfo {author} {\bibfnamefont {S.}~\bibnamefont {Pappalardi}}\ and\ \bibinfo {author} {\bibfnamefont {J.}~\bibnamefont {Kurchan}},\ }\href {https://doi.org/10.3390/e25020246} {\bibfield  {journal} {\bibinfo  {journal} {Entropy}\ }\textbf {\bibinfo {volume} {25}},\ \bibinfo {pages} {246} (\bibinfo {year} {2023})}\BibitemShut {NoStop}%
\bibitem [{\citenamefont {Lerose}\ and\ \citenamefont {Pappalardi}(2020)}]{pap3}%
  \BibitemOpen
  \bibfield  {author} {\bibinfo {author} {\bibfnamefont {A.}~\bibnamefont {Lerose}}\ and\ \bibinfo {author} {\bibfnamefont {S.}~\bibnamefont {Pappalardi}},\ }\href {https://doi.org/10.1103/PhysRevA.102.032404} {\bibfield  {journal} {\bibinfo  {journal} {Phys. Rev. A}\ }\textbf {\bibinfo {volume} {102}},\ \bibinfo {pages} {032404} (\bibinfo {year} {2020})}\BibitemShut {NoStop}%
\bibitem [{\citenamefont {Micklitz}\ \emph {et~al.}(2019)\citenamefont {Micklitz}, \citenamefont {Monteiro},\ and\ \citenamefont {Altland}}]{micklitz2019nonergodic}%
  \BibitemOpen
  \bibfield  {author} {\bibinfo {author} {\bibfnamefont {T.}~\bibnamefont {Micklitz}}, \bibinfo {author} {\bibfnamefont {F.}~\bibnamefont {Monteiro}},\ and\ \bibinfo {author} {\bibfnamefont {A.}~\bibnamefont {Altland}},\ }\href {https://doi.org/10.1103/PhysRevLett.123.125701} {\bibfield  {journal} {\bibinfo  {journal} {Phys. Rev. Lett.}\ }\textbf {\bibinfo {volume} {123}},\ \bibinfo {pages} {125701} (\bibinfo {year} {2019})}\BibitemShut {NoStop}%
\bibitem [{\citenamefont {Monteiro}\ \emph {et~al.}(2021)\citenamefont {Monteiro}, \citenamefont {Micklitz}, \citenamefont {Tezuka},\ and\ \citenamefont {Altland}}]{monteiro2021minimal}%
  \BibitemOpen
  \bibfield  {author} {\bibinfo {author} {\bibfnamefont {F.}~\bibnamefont {Monteiro}}, \bibinfo {author} {\bibfnamefont {T.}~\bibnamefont {Micklitz}}, \bibinfo {author} {\bibfnamefont {M.}~\bibnamefont {Tezuka}},\ and\ \bibinfo {author} {\bibfnamefont {A.}~\bibnamefont {Altland}},\ }\href {https://doi.org/10.1103/PhysRevResearch.3.013023} {\bibfield  {journal} {\bibinfo  {journal} {Phys. Rev. Res.}\ }\textbf {\bibinfo {volume} {3}},\ \bibinfo {pages} {013023} (\bibinfo {year} {2021})}\BibitemShut {NoStop}%
\bibitem [{\citenamefont {Dieplinger}\ \emph {et~al.}(2021)\citenamefont {Dieplinger}, \citenamefont {Bera},\ and\ \citenamefont {Evers}}]{dieplinger2021ansykinspired}%
  \BibitemOpen
  \bibfield  {author} {\bibinfo {author} {\bibfnamefont {J.}~\bibnamefont {Dieplinger}}, \bibinfo {author} {\bibfnamefont {S.}~\bibnamefont {Bera}},\ and\ \bibinfo {author} {\bibfnamefont {F.}~\bibnamefont {Evers}},\ }\href {https://doi.org/https://doi.org/10.1016/j.aop.2021.168503} {\bibfield  {journal} {\bibinfo  {journal} {Ann. Phys.}\ }\textbf {\bibinfo {volume} {435}},\ \bibinfo {pages} {168503} (\bibinfo {year} {2021})}\BibitemShut {NoStop}%
\bibitem [{\citenamefont {Leone}\ \emph {et~al.}(2023{\natexlab{b}})\citenamefont {Leone}, \citenamefont {Oliviero},\ and\ \citenamefont {Hamma}}]{leone2023nonstabilizerness}%
  \BibitemOpen
  \bibfield  {author} {\bibinfo {author} {\bibfnamefont {L.}~\bibnamefont {Leone}}, \bibinfo {author} {\bibfnamefont {S.~F.~E.}\ \bibnamefont {Oliviero}},\ and\ \bibinfo {author} {\bibfnamefont {A.}~\bibnamefont {Hamma}},\ }\href {https://doi.org/10.1103/PhysRevA.107.022429} {\bibfield  {journal} {\bibinfo  {journal} {Phys. Rev. A}\ }\textbf {\bibinfo {volume} {107}},\ \bibinfo {pages} {022429} (\bibinfo {year} {2023}{\natexlab{b}})}\BibitemShut {NoStop}%
\bibitem [{\citenamefont {Haug}\ \emph {et~al.}()\citenamefont {Haug}, \citenamefont {Lee},\ and\ \citenamefont {Kim}}]{haug2023efficient}%
  \BibitemOpen
  \bibfield  {author} {\bibinfo {author} {\bibfnamefont {T.}~\bibnamefont {Haug}}, \bibinfo {author} {\bibfnamefont {S.}~\bibnamefont {Lee}},\ and\ \bibinfo {author} {\bibfnamefont {M.~S.}\ \bibnamefont {Kim}},\ }\href@noop {} {}\Eprint {https://arxiv.org/abs/2305.19152} {arxiv:2305.19152} \BibitemShut {NoStop}%
\end{thebibliography}
\end{document}